\documentclass[aps,prb,twocolumn,superscriptaddress]{revtex4}
\usepackage{graphicx}
\usepackage{latexsym}
\usepackage{amssymb}
\usepackage{amsmath}
\usepackage{amsfonts}
\usepackage{upgreek}
\usepackage{bm}
\usepackage{multirow}
\usepackage{color}

\newcommand{\ket}[1]{|#1 \rangle}
\newcommand{\dd}{\mathrm{d}}
\newcommand{\ii}{\mathrm{i}}
\newcommand{\upc}{\mathrm{c}}
\newcommand{\U}{\mathrm{U}}
\newcommand{\SU}{\mathrm{SU}}
\renewcommand{\O}{\mathrm{O}}
\newcommand{\SO}{\mathrm{SO}}
\newcommand{\Sp}{\mathrm{Sp}}
\newcommand{\dsZ}{\mathbb{Z}}
\newcommand{\dsR}{\mathbb{R}}

\newcommand{\scL}{\mathcal{L}}
\newcommand{\scH}{\mathcal{H}}
\newcommand{\scK}{\mathcal{K}}
\newcommand{\scT}{\mathcal{T}}
\newcommand{\scS}{\mathcal{S}}
\newcommand{\Tr}{\operatorname{Tr}}
\newcommand{\sgn}{\operatorname{sgn}}
\renewcommand{\Re}{\operatorname{Re}}
\renewcommand{\Im}{\operatorname{Im}}
\newcommand{\vect}[1]{{\bm{#1}}}

\newcommand{\mat}[1]{\left[\begin{matrix}#1\end{matrix}\right]}
\newcommand{\smat}[1]{\left[\begin{smallmatrix}#1\end{smallmatrix}\right]}
\newcommand{\eq}[1]{\begin{equation}#1\end{equation}}
\newcommand{\eqs}[1]{\begin{equation}\begin{split}#1\end{split}\end{equation}}
\newcommand{\eqnref}[1]{Eq.\,\eqref{#1}}
\newcommand{\figref}[1]{Fig.\,\ref{#1}}
\newcommand{\tabref}[1]{Tab.\,\ref{#1}}

\begin{document}

\title{From Bosonic Topological Transition to Symmetric Fermion Mass Generation}

\author{Yi-Zhuang You}

\author{Yin-Chen He}
\affiliation{Department of Physics, Harvard University, Cambridge, MA 02138, USA}

\author{Ashvin Vishwanath}
\affiliation{Department of Physics, Harvard University, Cambridge, MA 02138, USA}

\author{Cenke Xu}
\affiliation{Department of Physics, University of California,
Santa Barbara, CA 93106, USA}

\date{\today}
\begin{abstract}

The bosonic topological transition (BTT) is a quantum critical point between the bosonic symmetry protected topological phase and the trivial phase. In this work, we derive a description of this transition in terms of compact quantum electrodynamics (QED) with four fermion flavors ($N_f=4$). This allows us to describe the transition in a lattice model with the maximal microscopic symmetry: an internal $\SO(4)$ symmetry. Within a systematic renormalization group analysis, we identify the critical point with the desired $\O(4)$ emergent symmetry and all expected deformations. By lowering the microscopic symmetry we recover the previous $N_f=2$ non-compact QED description of the BTT. Finally, by merging two BTTs we recover a previously discussed theory of symmetric mass generation, as an $\SU(2)$ quantum chromodynamics-Higgs theory with $N_f=4$ flavors of $\SU(2)$ fundamental fermions and one $\SU(2)$ fundamental Higgs boson. This provides a consistency check on both theories. 

\end{abstract}
\maketitle

\section{Introduction}

Over the past decade, much progress has been made in understanding
topological phases,\cite{Wen:2016xr} especially the symmetry
protected topological (SPT)
phases.\cite{Gu:2009ai,Chen:2011vx,Schuch:2011uv,Pollmann:2012jw,Chen:2013dt}
Concepts and methods developed in the study of SPT phases also
help to deepen our understanding of some \emph{gapless}
states,\cite{Wang:2015hb,Metlitski:2016dz,Karch:2016la,Seiberg:2016ru,Wang:2017zp}
which exist either on the boundary of SPT states or as the quantum
critical points between different SPT phases. Within all these
novel quantum critical points, the \emph{bosonic topological transition}
(BTT)\cite{Lu:2012ve,Grover:2013dz,Slagle:2015lo,He:2016ss,You:2016nr,Yoshida:2016kc,Wu:2016cz,Tsui:2017om} between a prototype (2+1)D bosonic symmetry protected
topological (BSPT) state and the trivial state has
attracted considerable attentions both theoretically and
numerically. It is believed that this transition is described by a
$N_f = 2$ {\it non-compact} QED$_3$, and it can have an as large as
O(4) emergent symmetry, due to its
self-duality~\cite{Xu:2015tm,Mross:2016rw,Hsin:2016ix}. Also, it has
been shown recently that this theory is dual to the easy-plane
deconfined quantum critical
point.\cite{Karch:2016la,Potter:2016th,Wang:2017zp}

Another novel transition that has been discussed in recent
literatures, in both condensed matter and high energy physics
communities, is the \emph{symmetric mass generation} (SMG)
transition
\cite{Slagle:2015lo,BenTov:2015lh,Catterall:2016sw,Ayyar:2016fi,Catterall:2016nh,Ayyar:2016tg,Ayyar:2016ph,He:2016qy,Huffman:2017rg,You:2017hy}.
This SMG quantum critical point was observed in various lattice models, using very
different numerical techniques. In the condensed matter community, the SMG was found in a fermion model on a double layer
honeycomb lattice,\cite{Slagle:2015lo,He:2016ss,You:2016nr} which
we will review later. Despite its different lattice realization,
the essence of the SMG is that, eight flavors of (two-component)
Dirac fermions in (2+1)D can generate a gap through short range
interaction without acquiring a nonzero expectation value of any
fermion bilinear operator. This transition is novel and unexpected
in the sense that it is clearly beyond the standard Gross-Neveu
mechanism of spontaneous generation of Dirac fermion mass, which
always corresponds to condensing fermion bilinear
operators.

In this paper, we will present a unified framework that captures
both novel phase transitions mentioned above. We first show that the BTT between the prototype (2+1)D $\SO(4)$
BSPT phases can be understood as a deconfined quantum critical
point (DQCP),\cite{Senthil:2004wj,Motrunich:2004hh,Senthil:2004qm,Senthil:2006lz} where the low-energy bosonic collective modes (below
the physical fermion gap) are fractionalized into $N_f=4$ flavors
of fermionic partons $\psi$ (four Dirac fermions) coupled with a
compact $\U(1)$ gauge field. The construction is based on the idea
of realizing BSPT states from interacting fermionic SPT
states,\cite{You:2014ho,Bi:2016zi,You:2016nr} where the bosonic
(spin and charge) freedom $\bar{\psi}\vect{\Gamma}\psi$ can be
treated as fermion bilinear order parameters. The BTT theory can
be described by the following Lagrangian,
\eq{\scL_\text{BTT}=\bar{\psi}\big(\gamma\cdot(\partial-\ii
a)+m\sigma^z\big)\psi+\scL_\text{int}[\psi], \label{eq: BTT2}}
where $\scL_\text{int}[\psi]$ contains the short-range
four-fermion interaction that explicitly breaks the $\SU(4)$
fermion flavor symmetry of the $N_f=4$ QED$_3$ theory down to its
$\SO(4)$ subgroup. The BTT is driven by the fermionic parton mass
$m$. At the critical point ($m=0$), we perform a large-$N_f$
renormalization group (RG) analysis\cite{Kaveh:2005tt,Xu:2008rz,Xu:2008jc} to
show that the short-range interaction $\scL_\text{int}[\psi]$ can
become relevant at the quantum electrodynamics (QED) fixed point,
which drives the theory to a new stable fixed point with the
desired $\O(4)$ symmetry of the prototype BSPT state, and the only
relevant symmetry allowed perturbation is the fermion mass in
\eqnref{eq: BTT2}. Away from the critical point ($m\neq 0$), the
fermionic parton opens a gap and becomes a band insulator (coupled
to gauge field). The non-trivial (or trivial) BSPT phase
corresponds to placing the fermionic parton in the corresponding
topological (or trivial) band structure. In this sense, the BTT
can be understood as a gauged version of the fermionic parton SPT
transition.

To make connection with the BTT, we propose that the SMG is a DQCP
as well, where the physical fermion is fractionalized into bosonic
$\phi$ and fermionic $\psi$ partons coupled together by a
non-Abelian gauge field $a^a\tau^a$.~\footnote{We will work with
Hermitian gauge connections throughout this work.} It can be
described by the following Lagrangian,
\eqs{\scL_\text{SMG}=&\;\bar{\psi}\gamma\cdot(\partial-\ii a^a\tau^a)\psi+|(\partial-\ii a^a\tau^a)\phi|^2\\
&+r|\phi|^2+u|\phi|^4+\cdots.} For the specific bilayer honeycomb
model studied extensively, it turns out that the most natural
emergent gauge group is $\SU(2)$, and the gauge field couples to
\emph{four} $\SU(2)$ fundamental fermions $\psi$ (totally eight
two-component Dirac fermions) and \emph{one} $\SU(2)$ fundamental
boson $\phi$. So this SMG theory is an $\SU(2)$ quantum
chromodynamics (QCD) with Higgs field $\phi$, resembling the
Standard Model in some aspects. Driven by the boson mass $r$, if
the bosonic parton $\phi$ condenses ($r<0$), the $\SU(2)$ gauge
field will be gapped out completely through the Higgs mechanism,
such that the fermionic parton $\psi$ effectively becomes the
physical fermion, which describes the semimetal phase with eight
gapless Dirac fermions at low energy. If the bosonic parton $\phi$
is gapped ($r>0$), we are left with a QCD theory decoupled from
the Higgs field $\phi$. Its fate in the IR limit is not fully
understood yet. But we assume that it is unstable towards
confinement by considering a spontaneous generation of a $\SU(2)$
gauge triplet mass. This mass will gap out the fermionic parton
and Higgs the gauge group down to $\U(1)$. The remaining $\U(1)$
gauge field is compact and will then lead to the confined phase.
Thus all excitations in the theory are gapped out and the system
enters the featureless insulator phase. The similar mechanism was also sketched in Ref.\,\onlinecite{Witten:2016yb}. If the same mass were
introduced to the physical fermion, it must break some symmetry.
However, for the fermionic parton, this triplet mass is not gauge
invariant, so the symmetry can be repaired by gauge transformation
in the form of the projective symmetry group
(PSG).\cite{Wen:2002qr} In this way, the featureless Mott
insulator can preserve the full symmetry of the Dirac semimetal,
therefore the transition between them is indeed a \emph{symmetric}
generation of the fermion mass (or more precisely, the fermion
gap). The similar Higgs-confine dichotomy in the $\SU(2)$
QCD-Higgs theory is also discussed in
Ref.\,\onlinecite{Chatterjee:2017ww} recently in the context of
cuprates high-$T_c$ superconductor.

It turns out that both BTT and SMG transitions can be realized in the phase diagram of a single lattice model (see Sec.\,\ref{sec: Phase Diagram} for more detailed discusssion). The two transitions are actually closely related. We show that the BTT theory can be obtained from the SMG theory by
gapping out the bosonic parton and half of the fermionic partons
(four out of the eight Dirac fermions). The remaining half of the
fermionic partons in the SMG theory actually become the fermionic
partons in the BTT theory. This consistency check lends more confidence to the
proposed SMG mechanism, as the assumed $\SU(2)$ gauge triplet mass
generation also plays a crucial role in the connecting BTT to SMG.

\section{Bosonic Topological Transition}

\subsection{Field Theory of BTT}

The 2D prototype BSPT has an SO(4) symmetry, and it corresponds
to the disordered phase of the (2+1)D $\O(4)$ non-linear sigma model (NLSM) with a $\Theta$-term at $\Theta=2\pi$. The BTT between the
prototype BSPT and the trivial state should have a field theory
description with an explicit O(4) symmetry. One possible field
theory preserving the $\SO(4)= \left(
\SU(2)_\uparrow\times\SU(2)_\downarrow \right) /\dsZ_2$
microscopic symmetry is the compact quantum electrodynamics (QED)
with fermion flavor number $N_f=4$.  In the bilayer honeycomb
lattice model (to be introduced in \eqnref{eq: KMH} later) where this BTT is realized, the four fermion
flavors can be arranged into two spin sectors
\eq{\psi=(\psi_\uparrow,\psi_\downarrow)^\intercal=(\psi_{\uparrow1},\psi_{\uparrow2},\psi_{\downarrow1},\psi_{\downarrow2})^\intercal.}
Within each spin sector (labelled by
$\sigma=\uparrow,\downarrow$), the fermion field $\psi_\sigma$
transforms as the fundamental representation of $\SU(2)_\sigma$,
\eq{U_\sigma\in\SU(2)_\sigma: \psi_\sigma\to U_\sigma\psi_\sigma,}
so altogether the field $\psi$ contains four Dirac fermions
transforming as the $\SO(4)$ spinor.\footnote{In Ref.\,\onlinecite{Jian:2017qe}, the four flavors of Dirac fermions form a SO(4) vector instead of a spinor. Thus the analysis in this paper will be very different form Ref.\,\onlinecite{Jian:2017qe}.} The theory can be considered
as a parton construction for the prototype BSPT state with
$\SO(4)$ symmetry, where the bosonic degree of freedom $\vect{N}$ (the $\O(4)$ vector in NLSM) is fractionalized into the fermionic parton
$\psi_\sigma$ as \eq{\label{eq: N parton}\vect{N}=\bar{\psi}_\uparrow(\mu^0,\ii\mu^1,\ii\mu^2,\ii\mu^3)\psi_\downarrow+\text{h.c.},}
with an emergent $\U(1)$ gauge field $a$. The BTT can be described
by the following compact $N_f=4$ QED theory \eqs{\label{eq:
BTT}\scL_\text{BTT}=&\sum_{\sigma}\bar{\psi}_{\sigma}
\big(\gamma\cdot(\partial-\ii a-\ii A_\sigma^a\mu^a)+m(-)^\sigma\big)\psi_{\sigma}\\
&+\frac{\ii}{8\pi}(\mathsf{CS}[A_\uparrow]-\mathsf{CS}[A_\downarrow])+\scL_{\text{int}}[\psi],}
with short-range interactions $\scL_{\text{int}}[\psi]$ that
explicitly break the $\SU(4)$ flavor symmetry\footnote{Strictly
speaking, the flavor symmetry should be $\SU(4)/\dsZ_4$, but we
will suppress the level of exactness in the following.} down to
$\SO(4)$. The shorthand notation $\gamma\cdot D$ denotes
$\gamma^\mu D_\mu$, where the gamma matrices are chosen as
$(\gamma^0,\gamma^1,\gamma^2)=(\sigma^2,\sigma^1,\sigma^3)$ and
$\bar{\psi}_\sigma=\psi_\sigma^\dagger\gamma^0$. The fermionic
partons couple to the \emph{compact} $\U(1)$ gauge field $a$. The
probe fields $A_\sigma=A_\sigma^a\mu^a$
($\sigma=\uparrow,\downarrow$) are introduced to keep track of the
$\SU(2)_\sigma$ symmetries, with
$\mu^a$ ($a=1,2,3$) being $\SU(2)$ generators (i.e.\,Pauli matrices). The background $\SU(2)$ Chern-Simons
term of the probe field $A_\sigma$ originates from the UV
regularization of the Dirac fermion.

The BTT is driven by the fermionic parton mass $m$. For $m\neq 0$, integrating out the fermion field $\psi$ in \eqnref{eq: BTT}
generates the following response theory \eqs{\label{eq: SU(2) CS}\scL_A&=\frac{\ii\nu}{4\pi}(\mathsf{CS}[A_\uparrow]-\mathsf{CS}[A_\downarrow]),\\
\mathsf{CS}[A_\sigma]&=\Tr\big(A_\sigma\wedge\dd
A_\sigma-\tfrac{2\ii}{3}A_\sigma\wedge A_\sigma\wedge
A_\sigma\big),} where
$\mathsf{CS}[A_\sigma]$ is $\SU(2)$ Chern-Simons term for the symmetry probe field $A_\sigma=A_\sigma^a\mu^a$ (in terms of Hermitian gauge connections). The topological index $\nu$
is given by $\nu=\frac{1}{2}(1+\sgn m)$. So $m<0$
corresponds to the featureless Mott phase ($\nu=0$) and $m>0$
corresponds to the BSPT phase ($\nu=1$). Therefore the BTT
transition should happen at $m=0$.

On general grounds, any interaction that is gauge invariant and
respects all physical symmetries could appear in
$\scL_\text{int}[\psi]$. Because the fermionic parton only
respects the physical symmetry $\SO(4)$, the interactions will
explicitly break the $\SU(4)$ flavor symmetry down to its $\SO(4)$
subgroup (at least at UV). We need to show that such interactions
are relevant so that the $\SU(4)$ symmetry will not be restored as
an emergent symmetry at IR (this is because the other theory
describing this transition, the $N_f = 2$ non-compact QED$_3$, can
at most host O(4) symmetry). Following the approach in
Ref.\,\onlinecite{Kaveh:2005tt}, we perform a large-$N_f$
controlled one-loop RG analysis and find an IR fixed point at
finite coupling strength with an emergent
$\O(4)=\SO(4)\rtimes\dsZ_2$ symmetry (where the
$\dsZ_2:\psi_\uparrow\leftrightarrow\psi_\downarrow$
transformation exchanges spin sectors and corresponds to the
improper $\O(4)$ rotation). We will postpone the details of the RG
analysis to Sec.\,\ref{sec: RG} and briefly summarize our main
findings in the following. We found that the most relevant
interaction takes the form of \eq{\label{eq: L_int
g}\scL_\text{int}[\psi]=\frac{g_2}{4\Lambda}\sum_{\sigma}\epsilon_{ac}\epsilon_{bd}(\bar{\psi}_{\sigma
a}\psi_{\sigma b})(\bar{\psi}_{\sigma c}\psi_{\sigma d}),} which
corresponds to the pair-pair interaction of $\SU(2)_\sigma$
singlets. $\Lambda$ denotes the UV cut-off. The RG equation for
$g_2$ reads (see Sec.\,\ref{sec: RG} \eqnref{eq: RG})
\eq{\frac{\dd
g_2}{\dd\ell}=-\Big(1-\frac{16}{\pi^2}\Big)g_2-\frac{1}{6\pi^2}g_2^2,}
which has an stable fixed point at $g_2^*=6\pi^2(16/\pi^2-1)$. It
will  be verified in Sec.\,\ref{sec: RG} that the $\dsZ_2$
symmetry breaking four-fermion interactions are irrelevant around
the fixed point (though the fermion mass $m$ in \eqnref{eq: BTT} that drives the BTT is still a $\dsZ_2$ breaking relevant perturbation). So
the fixed point has the desired emergent $\O(4)$ symmetry.

\subsection{Renormalization Group Analysis}\label{sec: RG}

In this section, we present the renormalization group (RG)
analysis of the $\SO(4)$ invariant four-fermion interactions in
the (2+1)D QED theory following
Ref.\,\onlinecite{Kaveh:2005tt}. To control the RG
calculation, we consider the $1/N_f$ expansion, where $N_f$ is the
fermion flavor number of the QED theory. In the end, we want to
apply the result to the physically relevant $N_f=4$ case.

First, we need to introduce a systematic large-$N_f$
generalization of the QED theory. One option is to start from the
$\SO(4)= \left( \Sp(1)_{\uparrow}\times\Sp(1)_{\downarrow} \right)
/\dsZ_2$ symmetry (at $N_f=4$) and promote the symmetry to $\left(
\Sp(N)_\uparrow\times\Sp(N)_\downarrow \right)/\dsZ_2$  with
$N_f=4N$. Then the QED theory in \eqnref{eq: BTT} can be
generalized to \eq{\label{eq:
L_QED(4N)}\scL_\text{QED}=\sum_{\sigma}\bar{\psi}_{\sigma}\gamma\cdot(\partial-\ii
a-\ii A_\sigma^a\mu^a)\psi_{\sigma}+\scL_{\text{int}}[\psi],}
where $\psi_\sigma=(\psi_{\sigma 1},\psi_{\sigma
2},\cdots,\psi_{\sigma 2N})^\intercal$ is now a $2N$ component
complex fermion field in each spin $\sigma=\uparrow,\downarrow$
sector. $A_\sigma^a$ are the source fields to keep track of the
$\Sp(N)_\uparrow\times\Sp(N)_\downarrow$ symmetry and $\mu^a$
denote the $\Sp(N)$ generators. The background response of
$A_\sigma$ is omitted.

Without the interaction $\scL_\text{int}[\psi]$, the $N_f=4N$ QED
theory in \eqnref{eq: L_QED(4N)} has the $\SU(4N)$ flavor
symmetry. The interaction will break the symmetry down to
$\Sp(N)_\uparrow\times\Sp(N)_\downarrow$. We will first impose an
additional $\dsZ_2$ symmetry that exchanges the two spin sectors,
\eq{\dsZ_2:\psi_\uparrow\leftrightarrow\psi_\downarrow,} which
reduces, in the $N=1$ case, to the improper $\dsZ_2$
transformation of the $\O(4)$ vector. The effect of breaking the
$\dsZ_2$ symmetry will be analyzed later. There are altogether
four independent types of the
$\Sp(N)_\uparrow\times\Sp(N)_\downarrow\rtimes\dsZ_2$ symmetric
interactions \eq{\label{eq: L_int
4N}\scL_\text{int}=\frac{1}{N_f\Lambda}(g_1 V_1+g'_1 V'_1+g_2
V_2+g_3 V_3),} where $g_1$, $g'_1$, $g_2$, $g_3$ are dimensionless
coupling constants and
\eqs{V_1&=(\textstyle\sum_{\sigma}\bar{\psi}_{\sigma}\psi_{\sigma})^2,\\
V'_1&=(\textstyle\sum_{\sigma}\bar{\psi}_{\sigma}\gamma^\mu\psi_{\sigma})^2,\\
V_2&=\textstyle\sum_{\sigma}J_{ac}J_{bd}(\bar{\psi}_{\sigma a}\psi_{\sigma b})(\bar{\psi}_{\sigma c}\psi_{\sigma d}),\\
V_3&=\textstyle\sum_{\sigma}J_{ac}J_{bd}(\bar{\psi}_{\sigma
a}\psi_{\bar{\sigma} b})(\bar{\psi}_{\sigma c}\psi_{\bar{\sigma}
d}),} where $\sigma=\uparrow,\downarrow$ labels the spin and
$\bar{\sigma}$ denotes the opposite spin of $\sigma$. $J$ is the
metric of the $\Sp(N)$ symplectic structure, such that
$J^\intercal=-J$, $J^2=-1$ and $\forall a: \mu^{a\intercal} J +J
\mu^{a}=0$. $g_1$ and $g'_1$ are actually $\SU(4N)$ symmetric and
irrelevant even for small $N$, see for example
Ref.\,\onlinecite{Xu:2008rz}. We will focus on the $\SU(4N)$
breaking interactions $g_2$ and $g_3$, which are all in the form
of the $\Sp(N)$ pair-pair coupling.

Following the derivation in
Ref.\,\onlinecite{Kaveh:2005tt,Xu:2008rz,Xu:2008jc}, we obtain the RG equation
for $g_2$ and $g_3$ as
\eqs{\label{eq: RG}\frac{\dd}{\dd\ell}g_2&=-\Big(1-\frac{64}{\pi^2 N_f}\Big)g_2-\frac{1}{6\pi^2}(g_2^2+g_3^2),\\
\frac{\dd}{\dd\ell}g_3&=-\Big(1-\frac{64}{\pi^2
N_f}\Big)g_3-\frac{1}{3\pi^2}g_2g_3.} At the QED ($g_i=0$) fixed
point, their scaling dimensions are degenerated
\eq{\Delta=-1+\frac{64}{\pi^2 N_f}+\mathcal{O}(1/N_f^2).} Pushing
this result to $N_f=4$, we obtain $\Delta\approx 0.6>0$, meaning
that the $g_2$ and $g_3$ interactions are relevant, which will
drive the theory alway from the QED fixed point and break the
$\SU(4)$ flavor symmetry down to $\O(4)=\SO(4)\rtimes\dsZ_2$. The
$g_2$ interaction reduces to the interaction term in \eqnref{eq:
L_int g} in the $N_f=4$ case.

\begin{figure}[htbp]
\begin{center}
\includegraphics[width=0.24\textwidth]{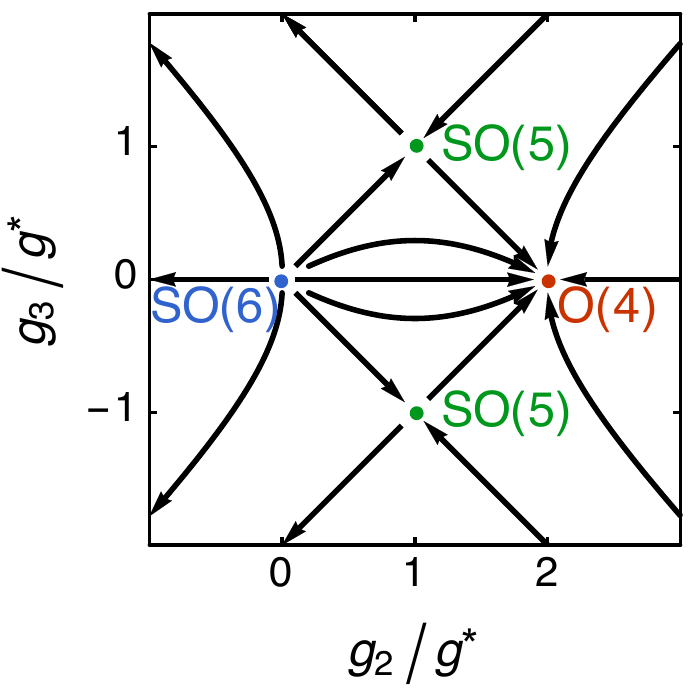}
\caption{RG flow diagram for $N_f=4$ ($N=1$) in the $\dsZ_2$
symmetric plane.} \label{fig: RG}
\end{center}
\end{figure}

We can track the RG flow away from the QED fixed point.
\figref{fig: RG} shows the RG flow diagram for the $N_f=4$ ($N=1$)
case. Several fixed points are found at finite couplings of the
order \eq{g^*=3\pi^2\Big(\frac{64}{\pi^2N_f}-1\Big).} First, there
are two $\SO(5)= \Sp(2)/\dsZ_2$ invariant fixed points at
$(g_2^*,g_3^*)=(g^*,\pm g^*)$. They are related by an $\SO(6)$
rotation. They both correspond to the $\SO(5)$ fixed point
discussed in Ref.\,\onlinecite{Xu:2008rz}, with different
$\SO(5)$ subgroups of the SO(6). Second, there is one stable
$\O(4)$ symmetric fixed point at $(g_2^*,g_3^*)=(2g^*,0)$.

Let us consider perturbing this fixed point by $\dsZ_2$ breaking
interactions $(g_4V_4+g_5V_5)/(N_f\Lambda)$ with
\eqs{V_4&=\textstyle\sum_{\sigma}(-)^\sigma J_{ac}J_{bd}(\bar{\psi}_{\sigma a}\psi_{\sigma b})(\bar{\psi}_{\sigma c}\psi_{\sigma d}),\\
V_5&=\textstyle\sum_{\sigma}\ii(-)^\sigma
J_{ac}J_{bd}(\bar{\psi}_{\sigma a}\psi_{\bar{\sigma}
b})(\bar{\psi}_{\sigma c}\psi_{\bar{\sigma} d}),} where
$(-)^\sigma=\pm1$ is the spin-dependent sign that discriminates
between $\sigma=\uparrow$ and $\downarrow$. It is found that both
$g_4$ and $g_5$ are irrelevant at the $\O(4)$ fixed point. 
Their RG equations are the same as $g_3$,
\eq{\frac{\dd}{\dd\ell}g_{i}=-\Big(1-\frac{64}{\pi^2
N_f}\Big)g_{i}-\frac{1}{3\pi^2}g_2g_{i}\quad (i=4,5).}
The scaling dimensions of $g_4$ and $g_5$ at the $\O(4)$ fixed point $g_2^*=2 g^*$ can be read off from the RG equation, which are given by $1-64/\pi^2 N_f \approx -0.6 <0$ and are irrelevant. So even if we start with the interaction that has only $\SO(4)$
symmetry (with $\dsZ_2$ broken terms $g_4$ and $g_5$), the QED
theory will flow to a stable fixed point with emergent
$\O(4)=\SO(4)\rtimes\dsZ_2$ symmetry. There is only one relevant
$\dsZ_2$ breaking perturbation at this fixed point, i.e.\,the mass
term $m(-)^\sigma\bar{\psi}_\sigma\psi_\sigma$ in \eqnref{eq:
BTT}, which drives the topological transition between the BSPT
phase and the featureless Mott phase.

One potential concern is that the fixed point may even have an
additional $\SO(2):\psi_\sigma\to
e^{(-)^\sigma\ii\theta}\psi_{\sigma}$ symmetry. However, this
$\SO(2)$ symmetry is broken by the monopole effect as pointed out
in Ref.\,\onlinecite{Wang:2017zp}. There it was argued that the
monopole operator $\mathcal{M}_{a}$ (which annihilates the $2\pi$
flux of the $a$ gauge field) transforms as an
$\SO(6)=\SU(4)/\dsZ_2$ vector. Required by the microscopic
$\SO(4)$ symmetry, four out of the six components of the $\SO(6)$
vector that transform under $\SO(4)$ are not allowed in the path
integral. So the $\SO(6)$ vector can only be aligned in the
remaining two component subspace, which rotates under $\SO(2)$.
This assigns the $\SO(2)$ symmetry charge to the monopole operator
$\mathcal{M}_{a}$. Due to the compactness of the $\U(1)$ gauge
field, the monopole term $\mathcal{M}_{a}+\text{h.c.}$ is allowed
in the Lagrangian. If we assume that the strength of the monopole
term remains finite at the $\O(4)$ fixed point, it will break the
above mentioned $\SO(2)$ symmetry completely. Therefore the $\O(4)$
RG fixed point will not have the additional $\SO(2)$ symmetry as
suspected. The scaling dimension of the monopole operator at
this $\O(4)$ fixed point requires further analysis. An alternative route to approach this $\O(4)$ fixed point is to start with the compact $N_f=4$ QED without the short-range parton interaction, the monopole argument in Ref.\,\onlinecite{Wang:2017zp} suggests that the theory describes an $\SO(5)$ DQCP fixed point. Upon the $\SO(5)$ to $\O(4)$ symmetry breaking anisotropy (corresponding to adding the parton interaction to the QED theory), the $\SO(5)$ fixed point becomes unstable and flows to the $\O(4)$ fixed point. Both understandings are consistent with our proposal that the $\O(4)$ fixed point can be described by the compact $N_f=4$ QED theory with parton interaction.

We would also like to mention that the $\SO(4)$ symmetry breaking
interactions $(g_6 V_6+g'_6 V'_6)/(N_f\Lambda)$, in the $(1,1)$
representation of $\SU(2)_\uparrow\times\SU(2)_\downarrow$, is
given by
\eqs{\label{eq: V6}V_6&=(\bar{\psi}_\uparrow\mu^3\psi_\uparrow) (\bar{\psi}_\downarrow\mu^3\psi_\downarrow),\\
V'_6&=(\bar{\psi}_\uparrow\gamma^\mu\mu^3\psi_\uparrow)
(\bar{\psi}_\downarrow\gamma^\mu\mu^3\psi_\downarrow),} where
$\mu^3$ is one of the $\SU(2)=\Sp(1)$ generators (so that
$J\mu^{3\intercal}J=\mu^{3}$). The RG flow equation of $g_6$ and
$g'_6$ around the $\O(4)$ fixed point is given by
\eqs{\label{eq: g6 RG}\frac{\dd}{\dd\ell}g_6&=-\Big(1-\frac{128}{3\pi^2N_f}-\frac{6g_2^*}{\pi^2N_f}\Big)g_6+\frac{64}{\pi^2N_f}g'_6,\\
\frac{\dd}{\dd\ell}g'_6&=-\Big(1+\frac{2g_2^*}{3\pi^2N_f}\Big)g'_6+\frac{64}{3\pi^2N_f}g_6,}
where $g_2^*=2g^*$. One can show there is a relevant channel along
$(g_6,g'_6)\propto(1,0.07)$, dominated by the $g_6V_6$
interaction. This interaction will drive the spontaneous
generation of the bilinear masses
$m_\sigma\bar{\psi}_\sigma\mu^3\psi_\sigma$. Depending on the sign
of $g_6$, either one of the $m_\uparrow=\pm m_\downarrow$ choices
will be favored, which leads to different spontaneous symmetry
breaking (SSB) phases that will be elaborated in Sec.\,\ref{sec:
SSB}.

In conclusion, the RG analysis indicates that the interacting
compact $N_f=4$ QED theory in \eqnref{eq: BTT} has a stable
$\O(4)$ fixed point, which has the emergent
$\O(4)=\SO(4)\rtimes\dsZ_2$ symmetry with only one $\dsZ_2$
breaking relevant perturbation and one $\SO(4)$ breaking relevant
perturbation in the $(1,1)$ representation. These properties are
all consistent with the known properties of BTT in other versions
of field theories.\cite{Xu:2015tm,Wang:2017zp} So we propose that
the $\SO(4)$ symmetric BTT can be equally described by the
interacting compact $N_f=4$ QED theory. In the following, we will
make connections to another field theory description of BTT with a
lower microscopic symmetry, which allows us to access the adjacent
SSB phases.

\subsection{Instability of BTT to SSB Phases}\label{sec: SSB}

In a series of recent theoretical\cite{Cheng:2016ky,Wang:2017zp}
and numerical\cite{Qin:2017os,Zhang:2017hl} works, it was pointed
out that the $\O(4)$ fixed point describes both the critical point
of $\SO(4)$ BTT and the deconfined quantum critical point
(DQCP)\cite{Senthil:2004wj,Motrunich:2004hh,Senthil:2004qm} of the
easy-plane N\'eel to valence bond solid (VBS) transition. The
N\'eel-VBS transition is driven by a relevant perturbation which
is in the $(1,1)$ representation of the $\SO(4)$ symmetry. However, in the context of our lattice model (to be reviewed later in Sec.\,\ref{sec: lattice}), the physical meaning of the $\O(4)$
vector $\vect{N}=(N_0,N_1,N_2,N_3)$ is interpreted differently. 
The N\'eel and the VBS order are interpreted as
the \emph{in-plane} spin density wave (SDW) order
$S^{+}\sim N_0+\ii N_3$ and the superconducting
(SC) order $\Delta^\dagger\sim N_2+\ii N_1$ respectively.
As analyzed in Sec.\,\ref{sec: RG} previously, the symmetry breaking perturbation in the $(1,1)$ representation corresponds to the following fermionic parton interaction in the $N_f=4$ QED theory,
\eq{\label{eq:
g6}\scL_\text{int}=\frac{g_6}{4\Lambda}(\bar{\psi}_{\uparrow}\mu^3\psi_{\uparrow})(\bar{\psi}_{\downarrow}\mu^3\psi_{\downarrow}).}
In terms of the $\O(4)$ order parameters, it can be also written as $g_6(N_1^2+N_2^2-N_0^2-N_3^2)$. This interaction discriminates between the SDW and the SC order. On the mean-field level, one could already see that $g_6>0$ (or $g_6<0$) favors the SDW (or SC) order.

With this interaction, the $\SO(4)$ symmetry is explicitly broken down
to its
$(\U(1)_\uparrow\times\U(1)_\downarrow)\rtimes\dsZ_2^{\updownarrow}$
subgroup, \eqs{\label{eq: UUZ
latt}\U(1)_\uparrow\times\U(1)_\downarrow&:
S^{+}\to e^{\ii(\theta_\uparrow-\theta_\downarrow)}S^{+},\\
&\hspace{9pt}\Delta^\dagger\to e^{-\ii(\theta_\uparrow+\theta_\downarrow)}\Delta^\dagger,\\
\dsZ_2^{\updownarrow}&:N_1\to-N_1,N_3\to-N_3,\\}
where $\U(1)_\uparrow\times\U(1)_\downarrow$ is a combination of the spin
and the charge $\U(1)$ symmetries and the
$\dsZ_2^{\updownarrow}$ conjugates both $\U(1)_\uparrow$ and
$\U(1)_\downarrow$ charges. According to the fractionalization scheme in \eqnref{eq: N parton}, the symmetry action on the fermionic partons are given by
\eqs{\label{eq: UUZ QED}\U(1)_{\uparrow}&:\psi_{\uparrow}\to e^{\frac{\ii}{2}\theta_\uparrow\mu^3}\psi_{\uparrow},\\
\U(1)_{\downarrow}&:\psi_{\downarrow}\to e^{\frac{\ii}{2}\theta_\downarrow\mu^3}\psi_{\downarrow},\\
\dsZ_2^{\updownarrow}&:\psi_{\sigma}\to
\ii\mu^2\psi_{\sigma}.}
The RG analysis in \eqnref{eq: g6 RG} shows that the $g_6$ interaction is relevant at the $\O(4)$ fixed point, meaning that under the explicit symmetry breaking to $(\U(1)_\uparrow\times\U(1)_\downarrow)\rtimes\dsZ_2^{\updownarrow}$, the $\SO(4)$ BTT critical point is unstable towards the SDW or the SC phase that further breaks the 
$\U(1)_\uparrow\times\U(1)_\downarrow$ symmetry spontaneously. These \emph{spontaneous symmetry breaking} (SSB) phases will set in between the BSPT phase and the featureless
Mott phase, splitting the BTT into two XY transitions. In the field theory, as $g_6$ flows to the strong coupling
limit, the interaction will drive the spontaneous generation of
the mass terms $m_\sigma\bar{\psi}_\sigma\mu^3\psi_\sigma$ (for
both $\sigma=\uparrow,\downarrow$). Depending on the sign of
$g_6$, the interaction will favor one of the $m_\uparrow\sim\pm
m_\downarrow$ choices, which corresponds to one of the $\U(1)$ SSB phases.

To analyze the effect of $m_\sigma$ mass terms in more details,
let us included them in the $N_f=4$ QED theory given in
\eqnref{eq: BTT} \eqs{\label{eq: BTT-DQCP}
\scL_{N_f=4}=&\sum_{\sigma}\bar{\psi}_\sigma\big(\gamma\cdot(\partial-\ii a-\ii A_\sigma^3\mu^3)+m(-)^\sigma\\
&\hspace{32pt}+m_\sigma\mu^3\big)\psi_\sigma+\scL_\text{bg}[A]+\scL_\text{int}[\psi].}
The $m_\sigma$ masses explicitly lowers the microscopic symmetry
from $\SO(4)$ to $\U(1)_\uparrow\times\U(1)_\downarrow$.
Correspondingly, the symmetry probe fields are reduced from the
non-Abelian field $A_\sigma^a\mu^a$ to the Abelian field
$A_\sigma^3\mu^3$, compared to \eqnref{eq: BTT}. The background
response is also reduced from the $\SU(2)$ Chern-Simons term in
\eqnref{eq: BTT} to its $\U(1)$ version
\eq{\scL_\text{bg}[A]=\sum_\sigma\frac{\ii}{4\pi}(-)^\sigma
A_\sigma^3\wedge\dd A_\sigma^3.} Let us first investigate the
possible phases that can be accessed by tuning the mass terms $m$
and $m_\sigma$. Suppose the fermionic parton is fully gapped by
these mass terms, the resulting Chern-Simons theory should take
the following form \eq{\label{eq:
CS}\scL_\text{CS}=\frac{\ii}{4\pi}K_{IJ}\mathcal{A}^I\wedge\dd\mathcal{A}^J,}
where $\mathcal{A}=(a,A_\uparrow^3,A_\downarrow^3)$ is a
collection of the gauge field $a$ and the symmetry probe fields
$A_\sigma^3$. The $K$ matrix in this basis is given by
\eqs{\label{eq: K mat}K=&\frac{1}{2}\sum_{\sigma,\mu}
\sgn{m_{\sigma\mu}}\mat{1 & (-)^\mu\delta_{\sigma\uparrow} & (-)^\mu\delta_{\sigma\downarrow}\\
(-)^\mu\delta_{\sigma\uparrow} & \delta_{\sigma\uparrow} & 0\\(-)^\mu\delta_{\sigma\downarrow} & 0 & \delta_{\sigma\downarrow}}\\
&+\mat{0&0&0\\0&1&0\\0&0&-1},} where
$m_{\sigma\mu}=m(-)^\sigma+m_\sigma(-)^\mu$ and $\sgn
m_{\sigma\mu}$ denotes the sign of $m_{\sigma\mu}$. The delta
symbol $\delta_{\sigma\sigma'}=1$ if $\sigma =\sigma'$ and
$\delta_{\sigma\sigma'}=0$ otherwise. The first term \eqnref{eq: K
mat} is obtained by integrating out the fermionic parton, and the
second term is from the background $\scL_\text{bg}$.

\begin{figure}[htbp]
\begin{center}
\includegraphics[width=0.4\textwidth]{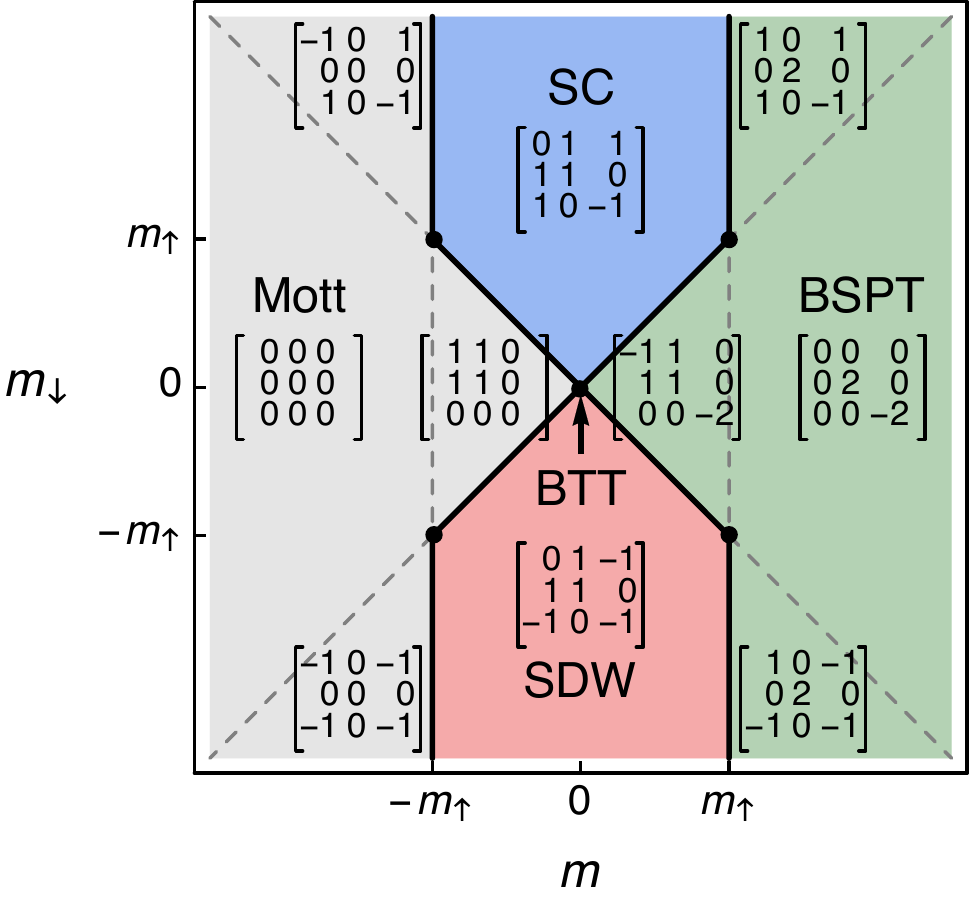}
\caption{Phase diagram of the model \eqnref{eq: BTT-DQCP} and $K$
matrices in different parameter regimes. The $K$ matrices are
given in the basis of
$\mathcal{A}=(a,A_\uparrow^3,A_\downarrow^3)$. We assume
$m_\uparrow>0$, and choose it as the mass scale. The solid lines
are physical phase transitions, while the dash lines are not.}
\label{fig: Kmat}
\end{center}
\end{figure}

There is a rich variety of $K$ matrices in the parameter space as
shown in \figref{fig: Kmat}. But in the end, there are only four
phases, since different $K$ matrices could describe the same
phase. For example, the following two $K$ matrices both correspond
to the BSPT phase
\eq{K_\text{BSPT}^\text{CS}=\smat{-1&1&0\\1&1&0\\0&0&-2}\sim
K_\text{BSPT}^\text{conf.}=\smat{0&0&0\\0&2&0\\0&0&-2}.} In both
cases, the gauge field $a$ is fully gapped, either due to the
Chern-Simons effect in $K_\text{BSPT}^\text{CS}$ or due to the
confinement in $K_\text{BSPT}^\text{conf.}$, leaving no
excitations at low energy. Their resulting response theories are
also identical. So there should be no phase transition between
them. Across this ``fake transition'' only one flavor of the
fermionic parton becomes gapless,
\eq{\bar{\psi}\gamma\cdot(\partial-\ii(a-A_\uparrow^3))\psi-\frac{\ii}{8\pi}a\wedge\dd
a+\frac{\ii}{4\pi}a\wedge \dd A_\uparrow^3+\cdots.} By redefining
$a\to a+A_\uparrow^3$, the Chern-Simons term $a\wedge \dd
A_\uparrow^3$ can be canceled exactly, leading to a \emph{compact}
$N_f=1$ QED theory with a level-1/2 $a\wedge \dd a$ term, which is completely
decoupled from all the symmetry probe fields $A_\sigma^3$.
Tuning the mass $\bar{\psi}\psi$ in this theory appears to drive a
``transition" between a confined phase of $a$ and another phase
with massive fermion coupled with a level-1 $a\wedge\dd a$ term. Both
sides correspond to trivial gapped phases of gauge invariant degrees of
freedom, thus this ``transition" should not exist. Similar
argument applies to other ``fake transitions'' (dashed lines) in
the phase diagram in \figref{fig: Kmat}. On the other hand,
the physical transitions (solid lines) in \figref{fig: Kmat} are
all described by \emph{non-compact} $N_f=1$ QED theories with
level-1/2 $a\wedge \dd a$ terms, which are dual to 3D XY
(Wilson-Fisher) transitions according to the fermion-boson
duality\cite{Potter:2016th,Seiberg:2016ru} as expected.  From the duality point of view, switching from the non-compact to the compact QED theory corresponds to explicitly breaking the $\U(1)$ symmetry that defines the XY transition in the dual theory,  such that the transition should be lifted along the dashed lines in \figref{fig: Kmat}. The
similar four-quadrant phase diagram among SPT and SSB phases was
also discussed in other (1+1)D\cite{Tsui:2017om} and
(2+1)D\cite{You:2016bj,Qin:2017os} systems.

Starting from the $\O(4)$ fixed point at the center of the phase
diagram, the mass $m$ drives the BTT along the horizontal
direction. In the BSPT (or featureless Mott) phase, the response
theory can be obtained by integrating out the gauge field $a$, which is found to be
\eq{\scL_{A}=\sum_\sigma\frac{\ii\nu}{2\pi}(-)^\sigma
A_\sigma^3\wedge\dd
A_\sigma^3=\frac{\ii\nu}{2\pi}A_\text{c}\wedge\dd A_\text{s},}
where $\nu=\frac{1}{2}(1+\sgn m)$. It is consistent with the
$\SO(4)$ version of the response theory in \eqnref{eq: SU(2) CS}.

On the other hand, $g_6$ interaction in \eqnref{eq: g6} drives the
$\O(4)$ fixed point into the SSB phases. As a relevant
interaction, a strong $g_6$ leads to the spontaneous generation of
the masses $m_\sigma$, which puts the system into the SSB phase.
From \eqnref{eq: UUZ QED}, one can see that the masses $m_\sigma$
are odd under $\dsZ_2^{\updownarrow}: m_{\sigma}\to
-m_{\sigma}$, so the $\dsZ_2^{\updownarrow}$ symmetry is
broken. Moreover, in the SSB phase, the Chern-Simons theory in
\eqnref{eq: CS} is reduced to
\eq{\scL_\text{CS}=\frac{\ii}{2\pi}\sum_{\sigma}(\sgn{m_{\sigma}})
A_\sigma^3\wedge\dd a+ \scL_\text{bg}[A].} For $m_\uparrow\sim\pm
m_\downarrow$, the gauge field $a$ is coupled to the probe field
$A^3_\uparrow\pm A^3_\downarrow$ by the Chern-Simons term, which
renders the gauge field $a$ non-compact. The gapless photon mode
of the $a$ field will be dual to the Goldstone mode in the SSB
phase. If $m_{\uparrow}$ and $m_{\downarrow}$ are of the opposite
(or same) sign, the gauge flux $\dd a$ will carry the spin (or
charge) quantum number and the gauge theory will describe the SDW
(or SC) phase.\cite{Cheng:2016ky} If we fix $m=0$ and tune $g_6$
interaction across zero, the $N_f=4$ QED theory will go through
the DQCP of the SDW-SC transition, which has the $\SO(4)$
microscopic symmetry and the $\O(4)$ emergent symmetry.

Finally, we would like to mention that there is a related but
different theory of BTT with a lower microscopic symmetry
$(\U(1)\rtimes\dsZ_2)\times\SU(2)$, where one of the
$\SU(2)_\sigma$ ($\sigma=\uparrow,\downarrow$) symmetry is broken
explicitly to its $\U(1)_\sigma\rtimes\dsZ_2^\sigma$ subgroup.
Without lost of generality, let us choose the microscopic symmetry
to be
$(\U(1)_\uparrow\rtimes\dsZ_2^\uparrow)\times\SU(2)_\downarrow$,
then the $m_\uparrow\bar{\psi}_\uparrow\mu^3\psi_\uparrow$ mass is allowed. Naively, a finite
$m_\uparrow$ seems to break the $\dsZ_2^\uparrow$ symmetry by picking one direction along the $\mu^3$-axis, but we
will see that the $\dsZ_2^\uparrow$ symmetry persists in the
low-energy effective theory as a particle-hole symmetry. Fixing a
finite mass $m_\uparrow>0$, two Dirac fermions $\psi_\uparrow$ in
the $N_f=4$ QED theory will be gapped, leaving an $N_f=2$ QED
thoery for the fermion $\psi_\downarrow$ at the
BTT.\cite{Grover:2013dz} The effective theory in \eqnref{eq:
BTT-DQCP} is thus reduced to \eqs{\label{eq: N=2 QED}
\scL_{N_f=2}&=\bar{\psi}_\downarrow\big(\gamma\cdot(\partial-\ii a-\ii A_\downarrow^a\mu^a)-m+m_\downarrow\mu^3\big)\psi_\downarrow\\
&\hspace{12pt}+\frac{\ii}{2\pi}A_\uparrow^3\wedge\dd
a+\scL_\text{bg}[A]+\scL_\text{int}[\psi].} The gauge field $a$ is
non-compact in this theory, and the conserved gauge flux $\dd a$
corresponds to the $\U(1)_\uparrow$ symmetry charge. The
$\dsZ_2^\uparrow:\psi_\downarrow\to\psi_\downarrow^\dagger,
a\to-a$ symmetry is realized as the particle-hole symmetry.

Driven by $m$ and $m_\downarrow$, all four phases in the phase
diagram \figref{fig: Kmat} can be realized within the framework of
the $N_f=2$ QED theory as
well.\cite{Xu:2015tm,Cheng:2016ky,Qin:2017os} They are separated
by quantum phase transitions. These phase transitions are of 3D XY
universality class, described by the non-compact $N_f=1$ QED
theory coupled to the ``level-1/2 Chern-Simons term''.\cite{Potter:2016th,Seiberg:2016ru} The four XY transition
lines join at the BTT multi-critical point, described by the
non-compact $N_f=2$ QED theory in \eqnref{eq: N=2 QED}, which also
has the emergent $\O(4)=\SO(4)\rtimes\dsZ_2$ symmetry at low
energy,\cite{Senthil:2006lz} where the improper
$\dsZ_2:\psi_\uparrow\leftrightarrow\psi_\downarrow$ symmetry is
realized as the
self-duality.\cite{Xu:2015tm,Mross:2016rw,Karch:2016la} The
self-dual $N_f=2$ QED is also dual to the non-compact CP$^1$
theory via the fermion-boson
duality.\cite{Potter:2016th,Seiberg:2016ru,Wang:2017zp} These dual
theories all describe the BTT multi-critical point, which has two
relevant perturbations: one leads to the transition between the
featureless Mott and the BSPT phases, and the other leads to the
transition between two SSB phases. The scenarios of the BTT under
different microscopic symmetries is concluded in \tabref{tab:
BTT}.

\begin{table}[htbp]
\caption{Effective descriptions of the BTT with different
microscopic symmetries.}
\begin{center}
\begin{tabular}{c|c}
microscopic symmetry & effective theory\\
\hline
$\SO(4)$ & critical, compact $N_f=4$ QED\\
\hline
$\begin{array}{c}\SU(2)_\uparrow\times(\U(1)_\downarrow\rtimes\dsZ_2^\downarrow)\\(\U(1)_\uparrow\rtimes\dsZ_2^{\uparrow})\times\SU(2)_\downarrow\end{array}$ & critical, non-compact $N_f=2$ QED\\
\hline
$\U(1)_\uparrow\times\U(1)_\downarrow\rtimes\dsZ_2^{\updownarrow}$ & not critical, SSB phases set in\\
\end{tabular}
\end{center}
\label{tab: BTT}
\end{table}

\subsection{Lattice Model and Symmetries}\label{sec: lattice}

\begin{figure}[htbp]
\begin{center}
\includegraphics[width=0.24\textwidth]{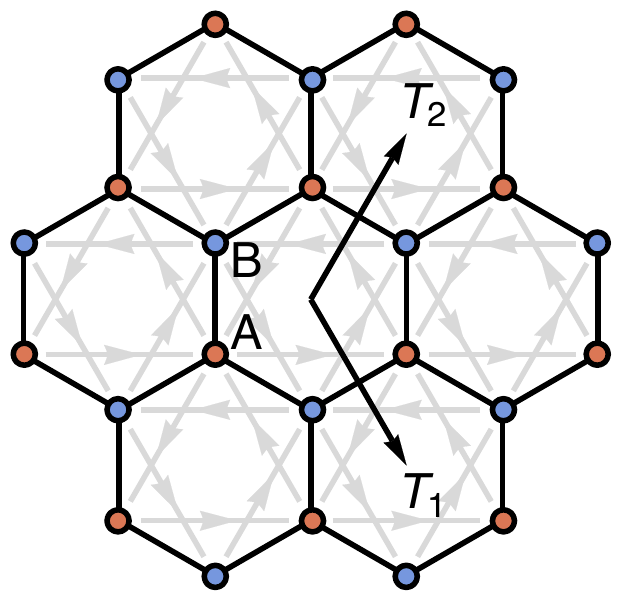}
\caption{Honeycomb lattice and space group symmetries. The lattice
can be partitioned into $A$ (red) and $B$ (blue) sublattices. The
sublattice sign $(-)^i$ is $+1$ on $A$ and $-1$ on $B$. The black
arrows mark the $T_{1,2}$ translation vectors. The background
arrow (light gray) shows Haldane's 2nd neighbor hopping direction
$\nu_{ij}$.} \label{fig: honeycomb}
\end{center}
\end{figure}

To be concrete, let us briefly review the lattice model that realizes the above mentioned $\SO(4)$ BTT. The
model is defined on the double layer honeycomb lattice as shown in
\figref{fig: honeycomb}, where the two sites from different layers
sit on top of each other (like the AA stacking bilayer
graphene\cite{Liu:2009ee}) and will be treated as a combined site.
On each site $i$ of the bilayer honeycomb lattice, there are four
complex fermion modes $c_{i\sigma\tau}$, where
$\sigma=\uparrow,\downarrow$ labels the \emph{spin} and $\tau=1,2$
labels the \emph{layer}. The fermion operators can be arranged
into the vector form $c_i=(c_{i\uparrow 1},c_{i\uparrow
2},c_{i\downarrow 1},c_{i\downarrow 2})^\intercal$. The model
Hamiltonian reads,\cite{You:2016nr}
\eqs{\label{eq: KMH}&H=H_\text{band}+H_\text{int},\\
&H_\text{band}=-t\sum_{\langle ij\rangle}c_{i}^\dagger c_{j} +
\lambda\sum_{\langle\!\langle i j\rangle\!\rangle}\ii \nu_{ij} c_i^\dagger \sigma^3 c_j+\text{h.c.},\\
&H_\text{int}=J\sum_{i}\big(\vect{S}_{i1}\cdot\vect{S}_{i2}+S_{i1}^zS_{i2}^z+\tfrac{1}{8}\rho_i^2\big).}
where $\vect{S}_{i\tau}=\frac{1}{2}c_{i\tau}^\dagger
\vect{\sigma}c_{i\tau}$ is the spin operator with Pauli matrices
$\vect{\sigma}=(\sigma^1,\sigma^2,\sigma^3)$ acting in the
\emph{spin} sector, and
$\rho_i=(\sum_{\sigma\tau}c_{i\sigma\tau}^\dagger
c_{i\sigma\tau}-2)$ is the on-site total charge density (measured
with respect to the half-filling). The Kane-Mele spin-orbit
coupling $\lambda$ is defined on the 2nd neighbor bonds with the
sign factor $\nu_{ij}$ being $+1$ ($-1$) for hopping along
(against) the bond direction specified in \figref{fig: honeycomb}.

The Hamiltonian in \eqnref{eq: KMH} has a pretty high symmetry
$\SO(4)\times\SO(3)_\ell$, where
$\SO(4)=(\SU(2)_\uparrow\times\SU(2)_\downarrow)/\dsZ_2$ and
$\SO(3)_\ell=\SU(2)_\ell/\dsZ_2$. It is sometimes more convenient
to lift the symmetry to
$\SU(2)_\uparrow\times\SU(2)_\downarrow\times\SU(2)_\ell$, which
can be defined by first rearranging the fermion operators
$c_{i\sigma\tau}$ on each site into the following matrix form
\eq{\label{eq: def
C}C_{i\sigma}=\mat{1&\\&(-)^\sigma}\mat{c_{i\sigma1}&-c_{i\sigma2}^\dagger\\c_{i\sigma2}&c_{i\sigma1}^\dagger}\mat{1&\\&(-)^i},}
where $(-)^\sigma$ is a staggered sign between spins,
\eq{(-)^\sigma=\left\{\begin{array}{ll}
+1 & \text{if $\sigma=\;\uparrow$,}\\
-1 & \text{if $\sigma=\;\downarrow$,}
\end{array}\right.}
and $(-)^i$ is a staggered sign between sublattices (see \figref{fig: honeycomb}).
\eq{\label{eq: sublattice sign}(-)^i=\left\{\begin{array}{ll}
+1 & \text{if $i\in A$ sublattice,}\\
-1 & \text{if $i\in B$ sublattice.}
\end{array}\right.}
Then the $\SU(2)_\uparrow\times\SU(2)_\downarrow\times\SU(2)_\ell$
symmetry acts on the matrix-form fermion operator $C_{i\sigma}$
(respectively for $\sigma=\uparrow,\downarrow$) as follows
\eq{\label{eq: def SU(2)s}C_{i\sigma}\to V
C_{i\sigma}U_\sigma^\dagger,} for $U_\sigma\in\SU(2)_\sigma$ and
$V\in\SU(2)_\ell$. Another way to understand the symmetry is to
view the four complex fermion modes on each site as eight Majorana
fermion modes, which form the eight-dimensional real spinor
representation of an $\SO(7)$ group, in which the symmetry group
$\SO(4)\times\SO(3)_\ell$ can be naturally embedded.

The action of these symmetries is most transparent by writing down
the fermion bilinear operators that transform as vectors under
$\SO(4)\times\SO(3)_\ell$. To this purpose, we define the  $\O(4)$
vector $\vect{N}_i$ and the $\O(3)$ vector $\vect{M}_i$ in terms
of the matrix-form fermion $C_{i\sigma}$ introduced in \eqnref{eq:
def C}, \eqs{\label{eq: def NM}
\vect{N}_i&=(-)^i\Tr\big(C_{i\downarrow}^\dagger C_{i\uparrow}(\mu^0,\ii\mu^1,\ii\mu^2,\ii\mu^3)\big),\\
\vect{M}_i&=(-)^i\sum_{\sigma}(-)^\sigma \tfrac{1}{2}\Tr
(C_{i\sigma}^\dagger\vect{\tau}C_{i\sigma}).} where $\mu^a$ and
$\tau^a$ ($a=0,1,2,3$) are Pauli matrices acting in the
\emph{particle-hole} and the \emph{layer} sectors respectively.
Under the
$\SU(2)_\uparrow\times\SU(2)_\downarrow\times\SU(2)_\ell$ symmetry
action defined in \eqnref{eq: def SU(2)s}, $\vect{N}_i$ rotates as
the vector representation of
$\SO(4)=(\SU(2)_\uparrow\times\SU(2)_\downarrow)/\dsZ_2$ and
$\vect{M}_i$ rotates as the vector representation of
$\SO(3)_\ell=\SU(2)_\ell/\dsZ_2$. It is instructive to label the
operators by the spin quantum numbers
$(s_\uparrow,s_\downarrow,s_\ell)$ of the
$\SU(2)_\uparrow\times\SU(2)_\downarrow\times\SU(2)_\ell$
symmetry, as summarized in \tabref{tab: symmetry charge}.

\begin{table}[htbp]
\caption{The
$\SU(2)_\uparrow\times\SU(2)_\downarrow\times\SU(2)_\ell$ symmetry
charge $(s_\uparrow,s_\downarrow,s_\ell)$ of various operators.}
\begin{center}
\begin{tabular}{cc}
operator & symmetry charge\\
\hline
physical fermion $c_i$ & $(\frac{1}{2},0,\frac{1}{2})\oplus(0,\frac{1}{2},\frac{1}{2})$\\
$\O(4)$ vector $\vect{N}_i$ & $(\frac{1}{2},\frac{1}{2},0)$\\
$\O(3)$ vector $\vect{M}_i$ & $(0,0,1)$
\end{tabular}
\end{center}
\label{tab: symmetry charge}
\end{table}

It can be checked that the model Hamiltonian in \eqnref{eq: KMH}
respects the $\SO(4)\times\SO(3)_\ell$ symmetry. In particular,
the complicated-looking interaction $H_\text{int}$ is such chosen
to preserve the symmetry. To make the symmetry property manifest,
the interaction can be rewritten as \eq{\label{eq: Hint O(3)}
H_\text{int}=-\frac{J}{8}\sum_{i}\vect{M}_i\cdot\vect{M}_i,} which
is just the inner product of the $\SO(3)_\ell$ vector on each site
and is obviously symmetric. The Hamiltonian also preserves some
lattice symmetries and the time-reversal symmetry, but we will
defer the discussion of those discrete symmetries when needed.

\subsection{Phase Diagram}\label{sec: Phase Diagram}

\begin{figure}[htbp]
\begin{center}
\includegraphics[width=0.36\textwidth]{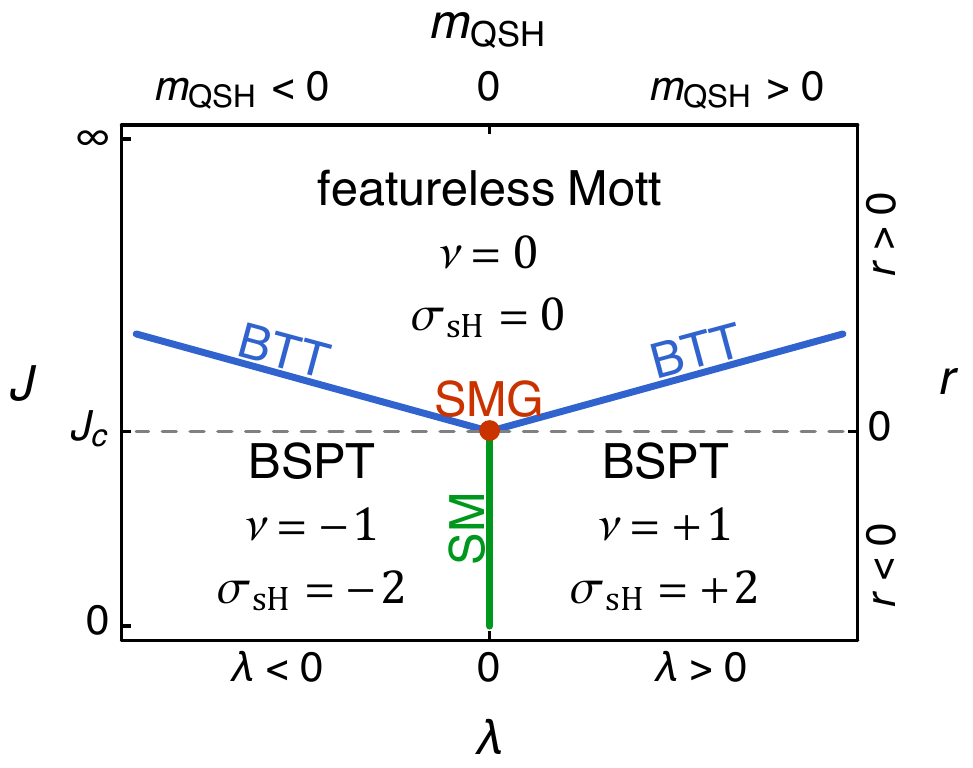}
\caption{Schematic phase diagram of both the bilayer honeycomb model
\eqnref{eq: KMH} tuned by $\lambda, J$ and the field theory \eqnref{eq: BSPT0} tuned by $m_\text{QSH}, r$. There is a featureless Mott phase and two
bosonic symmetry protected topological (BSPT) phases, labeled by the topological index $\nu$ or equivalently the quantum spin Hall conductance $\sigma_\text{sH}$. Different phases are separated by quantum phase transitions: a fermionic
transition (in green) corresponding to the Dirac semimetal (SM)
and two bosonic topological transitions (BTTs) (in blue). The
three transition lines meet at the symmetric mass generation (SMG)
tricritical point (in red). The dashed line is a ``faked transition'' in the field theory that does not correspond to any physical transition.} \label{fig: BSPT}
\end{center}
\end{figure}

The phase diagram of the lattice model 
\eqnref{eq: KMH} has been explored in several recent numerical works\cite{Slagle:2015lo,He:2016ss,Zhou:2016oa,Wu:2016cz,He:2016qy,Qin:2017os}. The phase diagram contains a featureless Mott phase and two (non-trivial) BSPT phases separated from each other by continuous quantum phase transitions. 
 In the free fermion limit ($J=0$), the
spin-orbit coupling $\lambda$ gaps out the fermion and drives the
system to a quantum spin Hall (QSH) insulator with spin Hall
conductance $\sigma_\text{sH}=2\sgn \lambda$, as illustrated in
\figref{fig: BSPT}. With weak interaction $J$, the QSH phase
becomes equivalent to the BSPT phase at low-energy, as the
fermionic edge modes are gapped out by the interaction and the
spin Hall current is now carried by collective bosonic edge
modes,\cite{You:2016nr,Wu:2016cz,Bi:2016zi} where the low-energy
bosonic freedom corresponds to the fermion bilinear order
parameter $\vect{N}$ defined in \eqnref{eq: def NM}. The BSPT
phases protected by the
$\SO(4)=(\SU(2)_\uparrow\times\SU(2)_\downarrow)/\dsZ_2$ symmetry are $\dsZ$ classified in (2+1)D. The topological index $\nu\in\dsZ$ can be defined as the level of the topological response theory in \eqnref{eq: SU(2) CS}. We label the BSPT phases by their topological index $\nu=\sgn \lambda$ in \figref{fig: BSPT}.

The topologically trivial phase ($\nu=0$) appears in the strong interacting limit $J\to\infty$. In this limit, the Hamiltonian is
dominated by $H_\text{int}$ in \eqnref{eq: KMH} and the ground
state of the system is simply a direct product of every on-site
ground state, \eq{\label{eq:
GS}\ket{\Psi}=\prod_i(c_{i\uparrow1}^\dagger c_{i\downarrow
2}^\dagger - c_{i\downarrow 1}^\dagger c_{i\uparrow
2}^\dagger)\ket{0}_c,} where $\ket{0}_c$ denotes the
$c_{i\sigma\tau}$ fermion vacuum state. From the on-site
interaction spectrum summarized in \tabref{tab: Hint spec}, one
can see that the ground state $\ket{\Psi}$ is unique,
fully-gapped, and $\SO(7)$ symmetric\footnote{When performing the
$\SO(7)$ transformation, one should keep in mind that the vacuum
state $\ket{0}_c$ may be rotated as well.} (which is also
$\SO(4)\times\SO(3)_\ell$ symmetric). Moreover, as a product
state, $\ket{\Psi}$ is topologically trivial by definition, hence
the topological index should vanish, i.e.~$\nu=0$. Above the
ground state, the excitation gap is of the order $J$, which can be
considered as a Mott gap. Small perturbations should not close the
gap, so the $\ket{\Psi}$ state actually represents a stable phase
in the large $J$ regime, which is the \emph{featureless Mott phase}\cite{Kimchi:2013fk,Jiang:2015qv,Kim:2016mz}. 

\begin{table}[htbp]
\caption{The spectrum of on-site interaction. $E_n$ and $d_n$ are
respectively the energy and the degeneracy of each level. The
ground state energy has been shifted to 0. All the eigenstates are
labeled by the representations of the $\SO(4)\times\SO(3)_\ell$
symmetry group (or of the larger $\SO(7)$ group).}
\begin{center}
\begin{tabular}{cccc}
$E_n$&$d_n$&representation&\\
$3J/2$&$\times 4$&$\SO(4)$ vector&\\
$9J/8$&$\times 8$&$\SO(7)$ spinor&\\
$J$&$\times 3$&$\SO(3)$ vector&\\
$0$&$\times 1$&$\SO(7)$ scalar&(ground state)
\end{tabular}
\end{center}
\label{tab: Hint spec}
\end{table}

Phases labeled by different topological indices ($\nu\in\dsZ$) must be
separated from each other by quantum phase transitions, as shown
in \figref{fig: BSPT}. The topological transition between the
$\nu=+1$ and $\nu=-1$ BSPT (or QSH) phases (driven by the
spin-orbit coupling $\lambda$ at weak interaction) is simply a
fermion gap closing transition. At this transition, the system
becomes a Dirac semimetal with eight gapless Dirac fermions at
low-energy. Since the short-range interaction $J$ is irrelevant
around the semimetal fixed point, the transition can be understood
within the free fermion band theory. A more exotic transition in
the phase diagram is the BTT, which is the transition between the featureless Mott
($\nu=0$) and the non-trivial BSPT ($\nu=\pm1$) phases. At the
transition, the physical fermions are gapped and only their
collective bosonic fluctuations become gapless, hence the
transition is \emph{bosonic}. As the SPT order changes across the
transition, the transition is also \emph{topological}. Several
recent numerical
simulations\cite{Slagle:2015lo,He:2016ss,Yoshida:2016kc} indicate
that the $\SO(4)$ symmetric BTT is a continuous transition. We propose that it can be described by the compact $N_f=4$ QED theory, analyzed in the previous discussion. 

The three phase boundaries in \figref{fig: BSPT} join at a tricritical point, known as the SMG critical point.\cite{You:2017hy} If we focus on the $\lambda=0$ axis, the
SMG can also be viewed as the transition that the eight Dirac
fermions in the semimetal phase are simultaneously gapped out by
the interaction $J$ without spontaneous symmetry breaking,
i.e.~without fermion bilinear condensation. A consistent theory
of SMG must be compatible with both the BTT and the semimetal
theory within the reach of perturbation. We have gained much
understanding of the BTT theory form the above discussion. To pin
down the SMG theory, we also need the input from the semimetal
side, which we will briefly review in the following.

\section{Symmetric Mass Generation}

\subsection{Semimetal: Field Theory and Symmetries}

The Dirac semimetal critical line refers to the fermionic
transition between the $\nu=\pm1$ BSPT phases (or more naturally
interpreted as weakly interacting QSH phases), which is along the
$\lambda=0$ axis with $J<J_c$ in the phase diagram \figref{fig:
BSPT} of the model \eqnref{eq: KMH}. In the semimetal phase,
interactions are irrelevant, and the effective theory simply
contains eight free Dirac fermions, or equivalently sixteen free
Majorana fermions $\upc\equiv(\Re c,\Im c)^\intercal$,
\eq{\label{eq:
SM}\scL_\text{SM}=\frac{1}{2}\sum_{Q,\sigma}\bar{\upc}
_{Q\sigma}\gamma\cdot(\partial-\ii A_\sigma^a\upmu^a-\ii
A_\ell^a\uptau^a)\upc_{Q\sigma}.} Hereinafter we will use the
upright letters (like $\upc$) for the real/Majorana fields, and
the italic letters (like $c$) for the complex/Dirac fields. The
Majorana fermion field $\upc_{Q\sigma}$ is labeled by the valley
index $Q=K_\pm$ and the spin index $\sigma=\uparrow,\downarrow$.
To simplify the representation of the
$\SU(2)_\uparrow\times\SU(2)_\downarrow\times\SU(2)_\ell$ symmetry
in the field theory, the valley modes
$\ket{K_\pm}=(\ket{K}\pm\ii\ket{K'})/\sqrt{2}$ are redefined as
combinations of the low-energy fermion modes from the $K$ and $K'$
points of the graphene Brillouin zone.\footnote{Otherwise, in the
original $(K,K')$ basis, the continuous symmetry transformations
will also involves rotations in the valley subspace, see
Appendix\,\ref{sec: Maj basis} for more details.} For each fixed
$Q$ and $\sigma$, the Majorana field $\upc_{Q\sigma}$ contains
eight real components: two for the Lorentz (sublattice $A,B$)
degrees of freedom, two for the layer ($\tau=1,2$), and two for
the particle-hole $(\Re c, \Im c)$. The adjoint Majorana fields
$\bar{\upc}_{Q\sigma}$ are defined as
$\bar{\upc}_{Q\sigma}=\upc_{Q\sigma}^\intercal\gamma^0$. The
external $\SU(2)$ gauge fields $A_\sigma$ and $A_\ell$ are
introduced to keep track of the $\SU(2)_\sigma$ and the
$\SU(2)_\ell$ symmetries respectively. Their charges (symmetry
group generators) are represented in the layer $\otimes$
particle-hole subspace as \eqs{\label{eq: symm charges}
A_\sigma=A_\sigma^a \upmu^a&: (\upmu^1,\upmu^2,\upmu^3)\equiv(\sigma^{23},\sigma^{21},\sigma^{02}),\\
A_\ell=A_\ell^a \uptau^a&:
(\uptau^1,\uptau^2,\uptau^3)\equiv(\sigma^{12},\sigma^{20},\sigma^{32}),}
where $\sigma^{ij}=\sigma^{i}\otimes\sigma^{j}$ and the 1st (2nd)
Pauli index belongs to the layer (particle-hole) subspace (see
Appendix\,\ref{sec: Maj basis} for derivation). Putting together
the valley, spin, layer and particle-hole degrees of freedom,
there are in total sixteen Majorana cones in the semimetal phase
of the double layer honeycomb model.

Besides the continuous on-site symmetry $\SO(4)\times\SO(3)_\ell$
defined in \eqnref{eq: def SU(2)s}, the lattice model also possess
the space group symmetry of the honeycomb lattice and several
anti-unitary symmetries. Among them, we will focus on the
translation symmetry and the chiral symmetry. There are two
linearly independent lattice translations, denoted by $T_1$ and
$T_2$ as shown in \figref{fig: honeycomb}. The chiral symmetry
$\dsZ_2^\scS$ (also known as the CT symmetry) is defined as
$\scS:c_i\to\scK(-)^ic_i^\dagger$, where $\scK$ denotes the
complex conjugation operator and $(-)^i$ is the sublattice sign
defined in \eqnref{eq: sublattice sign}. In the momentum space
(see Appendix\,\ref{sec: Maj basis}), both lattice translations
$T_{1}$ and $T_{2}$ are implemented as three-fold rotations in the
valley subspace, \eq{\label{eq: T12}
T_{1,2}:\mat{\upc_{K_+\sigma}\\\upc_{K_-\sigma}}\to\frac{1}{2}\mat{-1&\sqrt{3}\\-\sqrt{3}&-1}\mat{\upc_{K_+\sigma}\\\upc_{K_-\sigma}}.}
Note that $\ket{K_\pm}=(\ket{K}\pm\ii\ket{K'})/\sqrt{2}$ are
recombined valley modes, for which we define a valley sign factor
\eq{(-)^Q=\left\{\begin{array}{ll}
+1&\text{if $Q=K_+$,}\\
-1&\text{if $Q=K_-$.}
\end{array}\right.}
Then the chiral symmetry $\dsZ_2^\scS$ is implemented as
\eq{\label{eq: Z2S}\scS:\upc_{Q\sigma}\to\scK(-)^Q\ii\gamma^0
\upc_{\bar{Q}\sigma}.} The translation and the chiral symmetry
together is sufficient to rule out all the fermion bilinear
masses. This can be understood from an anomaly argument. The idea
is to construct an ``anomalous'' antiunitary symmetry
$\dsZ_2^\scT$ from the translation and chiral symmetries. Suppose
the translation symmetry can be enlarged from the $\dsZ_3$ valley
rotation to the $\U(1)$ rotation (as an emergent symmetry in the
field theory), by treating the conserved valley momentum as the
conserved ``charge'', then it can be used to generate a $\pi/2$
valley rotation
$R_{\pi/2}:\upc_{Q\sigma}\to(-)^{Q}\upc_{\bar{Q}\sigma}$, which
defines another antiunitary transformation $\scT=-R_{\pi/2}\scS$
from the chiral symmetry transformation $\scS$, \eq{\label{eq:
Z2T}\scT:\upc_{Q\sigma}\to\scK\ii\gamma^0 \upc_{Q\sigma}.} This
$\dsZ_2^\scT$ is not an on-site symmetry, so the field theory
\eqnref{eq: SM}  could behave anomalously under $\dsZ_2^\scT$. It
turns out that all Majorana fermions $\upc_{Q\sigma}$ transform in
the same way under $\dsZ_2^\scT$ with $\scT^2=-1$. This situation
is analogous to the (2+1)D Majorna fermions on the surface of the
(3+1)D class DIII topological superconductor (TSC) (e.g.\,the
$^3$He B phase).\cite{Schnyder:2008os,Ryu:2010fe,Ludwig:2016pt}
Without interaction, the class DIII TSC is $\dsZ$ classified in
(3+1)D, so the corresponding (2+1)D $\dsZ_2^\scT$-symmetric
Majorana fermions (as TSC surface states) are anomalous in the
non-interacting limit and can not be gapped out by fermion
bilinear masses without breaking the symmetry.

However, the $\dsZ\to\dsZ_{16}$ interaction reduced classification
of the class DIII
TSC\cite{Fidkowski:2013ww,Wang:2014lm,Metlitski:2014fp,You:2014ho,Gu:2015cy,Wang:2017ty}
implies that sixteen $\dsZ_2^\scT$-symmetric Majorana fermions are
actually anomaly free in the presence of interaction. So there
must be a way to gap out the sixteen Majorana fermions altogether
by interaction without breaking the translation and chiral
symmetry. Several field theory scenarios of how the sixteen Majorana fermions can be trivially gapped were proposed in an insightful work by Witten,\cite{Witten:2016yb} which generally require two separate transitions. In our lattice model \eqnref{eq:
KMH}, the fermions are gapped by the interlayer interaction $J$ though a single SMG transition, resulting in the
featureless Mott state directly. The SMG has been observed to be a continuous quantum phase transition in various different models\cite{Slagle:2015lo,Catterall:2016sw,Ayyar:2016fi,Catterall:2016nh,He:2016qy}. We will focus on our model \eqnref{eq: KMH}
throughout this work. Another model of SMG with a different
symmetry was discussed in Ref.\,\onlinecite{You:2017hy}, which
shares many common features. 

\subsection{Field Theory of SMG}

Now let us put all pieces of evidence together. The SMG is the
tricritical point in the phase diagram \figref{fig: BSPT} where
two BTT critical lines fuse into the semimetal critical line. Each BTT theory (the $N_f=4$
QED theory) contains four gauged Dirac fermions. Fusing two of
them together would result in eight gauged Dirac fermions, or
sixteen gauged Majorana fermions, which are fermionic partons that
transform under the $\SO(4)$ symmetry only. On the other hand, the
semimetal contains sixteen physical Majorana fermions, which
transform under the $\SO(4)\times\SO(3)_\ell$ symmetry. This
suggests that the fermionic parton should originate from the physical fermion by gauging
the $\SO(3)_\ell$ symmetry. Or more
precisely, we could consider fractionalizing the physical fermion
into the $\SO(4)$-charged fermionic parton and the
$\SO(3)_\ell$-charged bosonic parton, such that the gauge
structure emerges naturally.

To this end, we propose that the SMG could be described by the
following quantum chromodynamics (QCD) field theory (at $r=0$),
\eqs{\label{eq: SMG}
&\scL_\text{SMG}=\scL_\uppsi+\scL_\upphi,\\
&\scL_\uppsi=\frac{1}{2}\sum_{Q,\sigma}\bar{\uppsi}_{Q\sigma}\gamma\cdot(\partial-\ii a^a\uptau^a-\ii A_\sigma^a\upmu^a)\uppsi_{Q\sigma},\\
&\scL_\upphi=\frac{1}{2}\big((\partial-\ii a^a\uptau^a-\ii
A_\ell^a\upmu^a)\upphi\big)^2+\frac{r}{2}\upphi^2+\frac{u}{4}
\upphi^4,} which contains four $\SU(2)$ fundamental fermions
$\uppsi_{Q\sigma}$ (labeled by the valley $Q=K_\pm$ and the spin
$\sigma=\uparrow, \downarrow$ indices) and one $\SU(2)$
fundamental boson $\upphi$, both coupled to the internal $\SU(2)$
gauge field $a=a^a\uptau^a$. The external $\SU(2)$ gauge fields
$A_\uparrow=A_\uparrow^a\upmu^a$,
$A_\downarrow=A_\downarrow^a\upmu^a$ and $A_\ell=A_\ell^a\upmu^a$
are introduced to keep track of the
$\SU(2)_\uparrow\times\SU(2)_\downarrow\times\SU(2)_\ell$
symmetries of the physical fermion. The gauge/symmetry charges
$\uptau^a$ and $\upmu^a$ follow the same definition as in
\eqnref{eq: symm charges}. More specifically, $\upphi$ is a
four-component (gauge and particle-hole) real boson field, and
$\uppsi_{Q\sigma}$ is an eight-component (Lorentz, gauge and
particle-hole) Majorana fermion field (for each fixed valley $Q$
and spin $\sigma$). The matter fields are written in the
real/Majorana basis (indicated by their upright font), in order to
fully expose their symmetry properties.

The theory $\scL_\uppsi$ for the fermionic parton is very similar
to that of the physical fermion in the semimetal phase as
$\scL_\text{SM}$ in \eqnref{eq: SM} and the only difference is
that the symmetry field  $A_\ell$ in $\scL_\text{SM}$ is replaced
by the gauge field $a$. The $\SU(2)_\ell$ symmetry is now carried
by the bosonic parton $\upphi$ which also couples to the $\SU(2)$
gauge field $a$. The bosonic mass term $r\upphi^2\equiv r
\upphi^\intercal\upphi$ drives the theory between the semimetal
and featureless Mott phase across the SMG critical point. The $u
\upphi^4\equiv u(\upphi^\intercal\upphi)^2$ term in $\scL_\upphi$
simply remind us that the bosons are interacting. In general, all
symmetry-allowed and gauge-invariant interactions will be present
in the Lagrangian. We will not spell them all out explicitly.

\begin{figure}[hbtp]
\begin{center}
\includegraphics[width=0.42\textwidth]{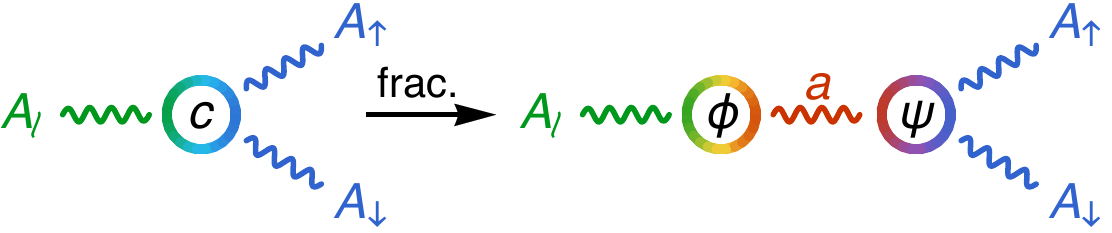}
\caption{The physical fermion $c$ carries three $\SU(2)$ symmetry
charges. Two of them, the $\SU(2)_\uparrow\times\SU(2)_\downarrow$
charges (in blue), are carried by the fermionic parton $\psi$; and
the remaining $\SU(2)_\ell$ charge (in green) is carried by the
bosonic parton $\phi$. Both partons carry the $\SU(2)$ gauge
charge (in red). The real fields $\upc$, $\upphi$ and $\uppsi$
will inherit the charge assignments.} \label{fig: frac}
\end{center}
\end{figure}

The fractionalization can be formulated on the lattice scale. To
simplify the presentation, we switch to the basis of complex
fermions and complex bosons. Let us introduce four complex
fermionic partons $\psi_{i\sigma\tau}$ and two complex bosonic
partons $\phi_{i\tau}$ on each site $i$, where
$\sigma=\uparrow,\downarrow$ and $\tau=1,2$ are the spin and the
layer indices. They are related to the real matter fields in the
field theory \eqnref{eq: SMG}, by
$\uppsi\equiv(\Re\psi,\Im\psi)^\intercal$ and
$\upphi\equiv(\Re\phi,\Im\phi)^\intercal$. We rearrange the
complex parton operators $\psi$ and $\phi$ into the matrix form
similar to \eqnref{eq: def C}, \eqs{\label{eq: def Psi Phi}
\Psi_{i\sigma}&=\mat{1&\\&(-)^\sigma}\mat{\psi_{i\sigma1}&-\psi_{i\sigma2}^\dagger\\\psi_{i\sigma2}&\psi_{i\sigma1}^\dagger}\mat{1&\\&(-)^i},\\
\Phi_{i}&=\mat{\phi_{i1}&-\phi_{i2}^\dagger\\\phi_{i2}&\phi_{i1}^\dagger}.}
Then the fractionalization of the physical fermion $C_{i\sigma}$
can be expressed as\cite{Hermele:2007zl}
\eq{\label{eq:
C=PhiPsi}C_{i\sigma}=\Phi_i^\dagger \Psi_{i\sigma}.} Under the
$\SU(2)_\uparrow\times\SU(2)_\downarrow\times\SU(2)_\ell$ symmetry
and $\SU(2)$ gauge, the parton operators transform as
\eq{\Phi_i\to G_i\Phi_iV^\dagger,\quad \Psi_{i\sigma}\to
G_i\Psi_{i\sigma}U_\sigma^\dagger,} for
$U_\sigma\in\SU(2)_\sigma$, $V\in\SU(2)_\ell$ and
$G_i\in\SU(2)_\text{gauge}$. Thus the physical fermion
$C_{i\sigma}$ in \eqnref{eq: C=PhiPsi} is gauge neutral and
transforms under the symmetries in the same way as defined in
\eqnref{eq: def SU(2)s}. The fractionalization scheme is depicted
in \figref{fig: frac}. The corresponding real fields $\upc$,
$\upphi$ and $\uppsi$ will inherit the similar fractionalization
relation $\upc_{Q\sigma}\sim\upphi\times\uppsi_{Q\sigma}$ from
\eqnref{eq: C=PhiPsi}.

For the convenience of later discussion, let us also define the
$\O(4)$ and $\O(3)$ vectors for the fermionic parton
$\uppsi_{Q\sigma}$, in analogy to that of the physical fermion
$\upc_{Q\sigma}$ following \eqnref{eq: def NM}. In terms of the
matrix-form operator in \eqnref{eq: def Psi Phi}, \eqs{\label{eq:
def NM for psi}
\vect{N}_i&=(-)^i\Tr\big(\Psi_{i\downarrow}^\dagger \Psi_{i\uparrow}(\sigma^0,\ii\sigma^1,\ii\sigma^2,\ii\sigma^3)\big),\\
\tilde{\vect{M}}_i&=(-)^i\sum_{\sigma}(-)^\sigma \tfrac{1}{2}\Tr
(\Psi_{i\sigma}^\dagger\vect{\tau}\Psi_{i\sigma}).} 
The parton $\O(3)$ vector $\tilde{\vect{M}}_i$ is different from $\vect{M}_i$ of the physical fermion defined in \eqnref{eq: def NM}, since $\tilde{\vect{M}}_i$ rotates under the $\SU(2)$ gauge transformation while $\vect{M}_i$ is gauge neutral and rotates under the $\SU(2)_\ell$ symmetry transformation. This difference is emphasized by the tilde in the notation of $\tilde{\vect{M}}_i$. On the other hand, the $\O(4)$ vector $\vect{N}_i$ still represents the physical bosonic order parameter as in \eqnref{eq: def NM}, which is also consistent with the fractionalization scheme in \eqnref{eq: N parton} in the BTT theory. The symmetry and gauge charges are summarize in \tabref{tab:
symmetry gauge charge}.

\begin{table}[htbp]
\caption{The
$\SU(2)_\uparrow\times\SU(2)_\downarrow\times\SU(2)_\ell$ symmetry
and the $\SU(2)$ gauge charges $(s_\uparrow,s_\downarrow,s_\ell;
s_\text{gauge})$ of various operators.}
\begin{center}
\begin{tabular}{cc}
field & (symmetry; gauge) charge\\
\hline
physical fermion $\upc_i$ & $(\frac{1}{2},0,\frac{1}{2};0)\oplus(0,\frac{1}{2},\frac{1}{2};0)$\\
fermionic parton $\uppsi_i$ & $(\frac{1}{2},0,0;\frac{1}{2})\oplus(0,\frac{1}{2},0;\frac{1}{2})$\\
bosonic parton $\upphi_i$ & $(0,0,\frac{1}{2};\frac{1}{2})$\\
$\psi$-parton $\O(4)$ vector $\vect{N}_i$ & $(\frac{1}{2},\frac{1}{2},0;0)$\\
$\psi$-parton $\O(3)$ vector $\tilde{\vect{M}}_i$ & $(0,0,0;1)$
\end{tabular}
\end{center}
\label{tab: symmetry gauge charge}
\end{table}

Besides the $\SO(4)\times\SO(3)_\ell$ continuous symmetry, let us
also briefly discuss the translation and the chiral symmetry of
the partons. From the on-site fractionalization scheme in
\eqnref{eq: C=PhiPsi}, it is obvious that the fermionic and the
bosonic partons will translate together in the same way as the
physical fermion.  What is non-trivial is the chiral symmetry
$\dsZ_2^\scS$ (and its derived symmetry $\dsZ_2^\scT$), whose
action on the partons is subject to $\SU(2)$ gauge
transformations. Such symmetry-gauge combined transformations form
the projective symmetry group (PSG)\cite{Wen:2002qr}, which
characterizes the symmetry fractionalization pattern of the
partons. As analyzed in Ref.\,\onlinecite{Witten:2016yb}, the PSG of the fermionic parton must be non-trivial in the SMG theory. In our context for example, one non-trivial choice of the PSG can be
\eq{\label{eq: PSG}\scT:\uppsi_{Q\sigma}\to\scK\ii\gamma^0
\ii\uptau^2\uppsi_{Q\sigma},\upphi\to\scK\ii\uptau^2\upphi, a\to
a.} In contrast to the $\scT^2=-1$ for the physical fermion
$\upc_{Q\sigma}$ in \eqnref{eq: Z2T}, the $\scT^2$ signature is
fractionalized to $\scT^2=+1$ on the fermionic parton
$\uppsi_{Q\sigma}$ and $\scT^2=-1$ on the bosonic parton $\upphi$.
This completely changes the anomaly classification for the
fermionic parton, as its symmetry class is shifted from DIII to
BDI. Given that the (3+1)D class BDI TSC has a trivial
classification, the Majorana fermions in parton QCD theory
$\scL_\uppsi$ (see \eqnref{eq: SMG}) is free from the
$\dsZ_2^\scT$ anomaly even on the fermion bilinear level. This
implies that a bilinear mass term for the fermionic parton is now
allowed by the $\dsZ_2^\scT$ PSG. This points out a plausible
route to get rid of the fermionic partons at low energy,\cite{Witten:2016yb} which
would eventually lead to the featureless Mott phase,
as to be elaborated in Sec.\,\ref{sec: SM}.

\subsection{From Semimetal to Featureless Mott Insulator}\label{sec: SM}

The bosonic parton mass $r$ in \eqnref{eq: SMG} is a relevant and
symmetric perturbation at the SMG critical point. The semimetal
phase can be accessed from the SMG critical point by condensing
the bosonic parton. When $r<0$, the bosonic parton condenses,
i.e.\,$\langle\upphi\rangle^2\neq 0$. Using the $\SU(2)$ gauge
freedom (which has three real gauge parameters), one can always
gauge the condensation direction to
$\langle\upphi\rangle=\upphi_0(1,0,0,0)^\intercal$, where
$\upphi_0\in\dsR$ is the condensation amplitude. Or equivalently,
$\langle\upphi\rangle$ can be written in the matrix form as
\eq{\langle\Phi\rangle=\upphi_0\mat{1&0\\0&1}.} As $\Phi\to G\Phi
V^\dagger$ for $V\in\SU(2)_\ell$ and $G\in\SU(2)_\text{gauge}$, to
keep $\langle\Phi\rangle$ invariant, we must have $G=V$, implying
that the $\SU(2)$ gauge field $a$ and the external $\SU(2)_\ell$
symmetry field $A_\ell$ are locked together. Therefore the
$\SU(2)_\ell$ symmetry remains unbroken. Its symmetry charge is
transferred to the fermionic parton $\uppsi_{Q\sigma}$, which is
now also equivalent to the physical fermion as
$\upc_{Q\sigma}\sim\langle\upphi\rangle\times\uppsi_{Q\sigma}$. So we have recovered the effective field theory of the semimetal phase in
\eqnref{eq: SM}. The translation and the chiral symmetry that
protects the gaplessness of the semimetal also remains unbroken.

The featureless Mott phase corresponds to the $r>0$ phase of
\eqnref{eq: SMG} where the bosonic parton is gapped out. At
low-energy (below the bosonic parton gap), we are left with the
QCD theory of the fermionic parton coupled to the $\SU(2)$ gauge
field, described by $\scL_\uppsi$ in \eqnref{eq: SMG}. The fate of
the QCD theory is not completely clear yet. Our conjecture is that an $\SU(2)$ gauge triplet bilinear mass is spontaneously generated
for the fermionic parton. We will supply this conjecture with more
evidence by making the connection to the BTT theory later. If we
accept such a spontaneous mass generation of the QCD theory, it
will gap out all the fermionic partons and also break the $\SU(2)$
gauge structure down to $\U(1)$. The compact $\U(1)$ gauge field
will then confine itself, since all matter fields have been gapped
out at this stage. Therefore we end up with a featureless ground
state with no gapless excitations, describing the featureless Mott
phase.

To be more precise, let us write down the effective field theory
for the featureless Mott phase. The $\SU(2)$ triplet mass
corresponds to the $\O(3)$ vector $\tilde{\vect{M}}$ of the
fermionic parton defined in \eqnref{eq: def NM for psi}. It is
only a matter of gauge choice to align
$\langle\tilde{\vect{M}}\rangle$ along the
$\langle\tilde{\vect{M}}\rangle\propto(0,0,1)$ direction. With
this gauge choice, the featureless Mott phase can be describe by
\eqs{\label{eq: Mott}
\scL_\text{Mott}&=\scL_\uppsi+m_\text{Mott}\tilde{M}^3\\
&=\frac{1}{2}\sum_{Q,\sigma}\bar{\uppsi}_{Q\sigma}\gamma\cdot(\partial-\ii a^3\uptau^3-\ii A_\sigma^a\upmu^a)\uppsi_{Q\sigma}\\
&\hspace{12pt}+\frac{1}{2}\sum_{Q,\sigma}m_\text{Mott}(-)^{Q+\sigma}\bar{\uppsi}_{Q\sigma}\ii\uptau^3\uppsi_{\bar{Q}\sigma}.}
If such a mass term $\vect{M}$ were introduced to the physical
fermion, it would break the $\SO(3)_\ell=\SU(2)_\ell/\dsZ_2$
symmetry and the chiral symmetry $\dsZ_2^\scS$ (or equivalently
$\dsZ_2^\scT$), because $\vect{M}$ transforms as a vector of the
$\SO(3)_\ell$ and is also odd under $\dsZ_2^\scT$
(i.e.\,$\scT:\vect{M}\to-\vect{M}$). However for the fermionic
parton, the vector $\tilde{\vect{M}}$ preserves all these
symmetries. First of all, the $\SO(3)_\ell$ is not broken because
the $\SO(3)_\ell$ symmetry charge has been carried away by the
bosonic parton, which is now in a gapped and disordered state. The
parton mass $\tilde{\vect{M}}$ only rotates under the $\SU(2)$
gauge transformation (as an $\SU(2)$ triplet). Then, because the
fermionic parton mass $\tilde{\vect{M}}$ is not gauge neural and
can be flipped by the gauge transformation, this leaves us rooms
to restore the ``broken" symmetry by the PSG. One can see that the
PSG transformation of $\dsZ_2^\scT$ in \eqnref{eq: PSG} is indeed
chosen to keep $\tilde{M}^3$ invariant. So the fermionic parton
mass $m_\text{Mott} \tilde{M}^3$ does not break any symmetry or
introduce any gauge anomaly.

With the $\SU(2)$ gauge triplet mass $m_\text{Mott}$, all the
fermions are gapped out. The $\SU(2)$ gauge field $a=a^a\tau^a$ is
reduced to a compact $\U(1)$ gauge field $a^3$ by the Higgs
mechanism. The expectation is that the compact $\U(1)$ gauge field
will get confined by the non-perturbative monopole effect.
However, there is the concern that the $\U(1)$ gauge flux might
carry some symmetry charges or projective representations, such
that the monopole operator would be forbidden by the symmetry and
the confinement could not occur. Here we show that this is not the
case.

First, we check the $\SO(4)$ symmetry. Integrating out the gapped
fermion $\uppsi_{Q\sigma}$ in \eqnref{eq: Mott}, no Chern-Simons
term is generated between the gauge field $a^3$ and the symmetry
probe fields $A_\sigma^a$, so the $\U(1)$ gauge flux $\dd a^3$
does not carry any $\SO(4)$ symmetry charge.

Next, we check the translation symmetry, by studying how the
$\U(1)$ gauge flux transforms under translation. To this purpose,
we calculate the on-site gauge charge
$\langle\uppsi_i^\intercal\uptau^3\uppsi_i\rangle$ and
$\langle\upphi_i^\intercal\uptau^3\upphi_i\rangle$ in the matter
field sector. It turns out that
$\langle\uppsi_i^\intercal\uptau^3\uppsi_i\rangle=\langle\upphi_i^\intercal\uptau^3\upphi_i\rangle=0$,
i.e.\,the matter field background is gauge neutral. So the $\U(1)$
gauge flux $\dd a^3$ does not see any background ``magnetic
field'' as it moves around on the lattice, meaning that the
translation symmetry is not fractionalized on  the $\U(1)$ gauge
flux ($T_1T_2T_1^{-1}T_2^{-1}=+1$). One may wonder why the
$\SU(2)$ triplet mean-field $\langle\tilde{\vect{M}}\rangle$ does
not lead to any gauge charge polarization in the fermionic sector.
This is because $\langle\tilde{\vect{M}}\rangle$ polarizes the
gauge charge oppositely in different spin
$\sigma=\uparrow,\downarrow$ sectors, as seen from \eqnref{eq:
Mott}, so there is no net gauge charge polarization on each site.
However, this also implies that the $\U(1)$ gauge flux $\dd a^3$
does carry the quantum number of a spin-dependent translation,
where $\uparrow$ and $\downarrow$ spins translate in opposite
directions (see Appendix\,\ref{sec: Maj basis} for derivation).
But this spin-dependent translation has been explicitly broken by
the interaction $H_\text{int}$ in the model Hamiltonian, so it
imposes no symmetry constraint on the monopole operator.

Finally, we check the chiral symmetry. The $\U(1)$ gauge flux is
reversed $\dd a^3\to-\dd a^3$ under the $\dsZ_2^\scS$ PSG, such
that the monopole operator $\mathcal{M}_{a^3}$ that creates the
gauge flux will be conjugated as
$\scS:\mathcal{M}_{a^3}\to\mathcal{M}_{a^3}^\dagger$. We also
verified numerically on the lattice that there is no sign/phase
change of the monopole operator $\mathcal{M}_{a^3}$ associated to
this conjugation. Therefore the monopole terms like
$\mathcal{M}_{a^3}+\mathcal{M}_{a^3}^\dagger$ are allowed by
symmetries in the Lagrangian. Such terms will drive the gauge
theory to the confined phase and gap out the $\U(1)$ photon from
the low-energy sector. In the end, all excitations in the theory
are gapped out, and we are left with a featureless Mott insulator.

\subsection{Accessing BSPT Phases}

In order to further check the consistency of the SMG theory, we
can perturb the SMG critical point by the Kane-Mele spin-obit
coupling $\lambda$, which breaks the chiral symmetry $\dsZ_2^\scS$
explicitly. As shown in the phase diagram \figref{fig: BSPT},
$\lambda$ is a relevant perturbation, which drives the system to
the BSPT phases. Within the framework of the SMG theory proposed
in \eqnref{eq: SMG}, the spin-orbit coupling $\lambda$ in the
lattice model \eqnref{eq: KMH} should correspond to a QSH mass
$m_\text{QSH}$ for the fermionic partons:
\eq{\label{eq: BSPT0}\scL_\text{BSPT}=\scL_\text{SMG}+\frac{1}{2}m_\text{QSH}\sum_{Q,\sigma}(-)^\sigma\bar{\uppsi}_{Q\sigma}\uppsi_{Q\sigma},}
because the $m_\text{QSH}$ mass term is a relevant perturbation
like $\lambda$ that also preserves the $\SO(4)\times\SO(3)_\ell$ and
the translation symmetry and breaks the chiral symmetry.

The topological response in the BSPT phase is easily found from
the fermionic parton sector,
\eqs{\label{eq: BSPT}\scL_{\text{BSPT}}=\frac{1}{2}\sum_{Q,\sigma}\bar{\uppsi}_{Q\sigma}\big(&\gamma\cdot(\partial-\ii a^a\uptau^a-\ii A_\sigma^a\upmu^a)\\
&+m_\text{QSH}(-)^\sigma\big)\uppsi_{Q\sigma}.} In the presence of
the $m_\text{QSH}$ mass, the fermionic parton $\uppsi_{Q\sigma}$
is fully gapped. Integrating them out, we obtain the Chern-Simon
term for the $\SU(2)_\sigma$ symmetry probe fields
$A_\sigma=A_\sigma^a\mu^a$\footnote{Here the symmetry charges
$\mu^a=\sigma^a$ are represented in the complex basis.} as
proposed in \eqnref{eq: SU(2) CS},
\eq{\scL_{A}=\frac{\ii\nu}{4\pi}(\mathsf{CS}[A_\uparrow]-\mathsf{CS}[A_\downarrow]).}
where the topological index is given by $\nu=\sgn m_\text{QSH}$.
No Chern-Simons term is generated for the gauge field $a$ or
between $a$ and $A_\sigma$. $\scL_{A}$ describes the response
theory of the $\SO(4)= \left(
\SU(2)_\uparrow\times\SU(2)_\downarrow \right)/\dsZ_2$ symmetric
BSPT state, with $\SU(2)_\uparrow$ and $\SU(2)_\downarrow$
currents running oppositely on the boundary.

What about the physics of the bosonic parton sector then? As the
QSH mass $m_\text{QSH}$ gaps out the fermionic parton and
generates the response theory $\scL_A$, the bosonic parton and the
$\SU(2)$ gauge field are left untouched. They are described by the
following theory \eq{\scL_\upphi=\frac{1}{2}\big((\partial-\ii
a^a\uptau^a-\ii
A_\ell^a\upmu^a)\upphi\big)^2+\frac{1}{2}r\upphi^2+\frac{1}{4}u
\upphi^4,} which is decoupled from the response theory $\scL_A$.
Despite the freedom to tune the parameter $r$, the theory
$\scL_\upphi$ has only one single phase (independent of $r$). When
$r<0$, the bosonic parton condenses, which Higgs out the $\SU(2)$
gauge field and attaches the $\SU(2)_\ell$ symmetry charge to the
fermionic parton, such that the physical fermion is restored (and
remains gapped). When $r>0$, the bosonic parton is gapped and the
fluctuating $\SU(2)$ gauge field will get confined, which binds
the partons into physics fermions in a gapped spectrum. In any
case, all excitations are gapped and response theory $\scL_A$ is
the same as that of the BSPT state. So there should be no physical
transition across the $r=0$ line (the dashed line in \figref{fig:
BSPT}) inside the BSPT phase. Both $\nu=\pm1$ BSPT
phases are accessible from the SMG critical point simply by
turning on the spin-orbit coupling for the physical fermions.

\subsection{Bridging SMG and BTT}

Now we are in the position to connect the SMG and BTT field
theories. Within the framework of the SMG theory, the competition between the featureless Mott phase and
the BSPT phase is just a matter of competing mass terms
$m_\text{Mott}$ and $m_\text{QSH}$ (on the $r>0$
side where the bosonic partons are gapped). Introducing both mass
terms to the fermionic parton (by merging \eqnref{eq: Mott} and
\eqnref{eq: BSPT}), the BTT theory can be derived as follows
\eqs{\label{eq: SMG BTT}\scL_\text{BTT}&=\frac{1}{2}\sum_{Q,\sigma}\bar{\uppsi}_{Q\sigma}\gamma\cdot(\partial-\ii a^3\uptau^3-\ii A_\sigma^a\upmu^a)\uppsi_{Q\sigma}\\
&+\frac{1}{2}\sum_{Q,\sigma}m_\text{Mott}(-)^{Q+\sigma}\bar{\uppsi}_{Q\sigma}\ii\uptau^3\uppsi_{\bar{Q}\sigma}\\
&+\frac{1}{2}\sum_{Q,\sigma}m_\text{QSH}(-)^\sigma\bar{\uppsi}_{Q\sigma}\uppsi_{Q\sigma}.}
The fermionic parton field $\uppsi_{Q\sigma}$ is still the same as in the SMG theory. But the $\SU(2)$ gauge structure is now broken down to $\U(1)$
in the presence of the gauge triplet mass $m_\text{Mott}$. The
remaining $\U(1)$ gauge field $a^3$ is compact. We may choose to
fix $m_\text{Mott}>0$ using the $\SU(2)$ gauge freedom in the SMG
theory.

Because the two masses $m_\text{Mott}$ and $m_\text{QSH}$ commute, they compete with each other to gap out the
fermionic parton in different manners. The BTT happens when the two
masses reaches a balance $|m_\text{Mott}|=|m_\text{QSH}|$, where
half of the fermionic partons in the theory will be gapped and the other half remain gapless.
The number of gapless fermions corresponds to eight Majorana or four Dirac,
matching the fermion falvor in the $N_f=4$ QED theory for the BTT. Obviously, the driving parameter of BTT is the difference
between $m_\text{Mott}$ and $m_\text{QSH}$, denoted as
\eq{m=m_\text{QSH}-m_\text{Mott}.} Herein we assume
$m_\text{QSH}>0$ by focusing on the BTT to the $\nu=+1$ BSPT
phase. In the vicinity of the BTT, we have $|m|\ll m_\text{QSH}$
and $m_\text{Mott}$, so there is a separation of the fermion mass
scale. The four massive Dirac fermions can be grouped into $\SU(2)_\uparrow$ and  $\SU(2)_\downarrow$ fundamentals with opposite masses $\pm(m_\text{QSH}+m_\text{Mott})$, providing the following ``level-1/2'' background
response \eq{\label{eq: BTT
bg}\scL_\text{bg}[A]=\frac{\ii}{8\pi}(\mathsf{CS}[A_\uparrow]-\mathsf{CS}[A_\downarrow]).}
The remaining four Dirac fermions are close to critical, which can
be describe by (see Appendix\,\ref{sec: Maj basis} for derivation)
\eq{\label{eq: BTT
psi}\sum_{\sigma}\bar{\psi}_{\sigma}\big(\gamma\cdot(\partial-\ii
a-\ii A_\sigma^a\mu^a)+m(-)^\sigma\big)\psi_\sigma,} where we have
switched back to the complex fermion basis (indicated by the italic font $\psi_\sigma$) and replace the compact $\U(1)$ gauge field
$a^3$ by $a$. Compared with the SMG theory, the absence of the valley index $Q$ in $\psi_{\sigma}$ reflects the fact that half of the fermions are
effectively removed from low-energy. In each spin $\sigma=\uparrow,\downarrow$ sector, the $\psi_\sigma$ fermion
transforms as the fundamental representation of $\SU(2)_\sigma$, so the 
$\SO(4)=\SU(2)_\uparrow\times\SU(2)_\downarrow/\dsZ_2$ symmetry can be implemented.
Putting together \eqnref{eq: BTT bg} and \eqnref{eq: BTT psi} and
including the $\SO(4)$ symmetry allowed interactions, we arrive
at the compact $N_f=4$ QED theory in \eqnref{eq: BTT} that was
proposed to describe the $\SO(4)$ symmetric BTT.

The connection to the BTT theory provides a piece of supportive
evidence for the existence of the $\SU(2)$ gauge triplet mass
$m_\text{Mott}$ in the Mott phase, which is one important
assumption in our explanation of the SMG. Another assumption we
made is that $m_\text{Mot}$ must be spontaneously generated once
the bosonic parton $\upphi$ is gapped, which we resorted to the
instability of the $N_f=4$ $\SU(2)$ QCD theory. If this assumption
is challenged, i.e.\,the bosonic parton gap $\sqrt{r}$ and the
fermionic parton gap $m_\text{Mott}$ do not open at the same
point, an intermediate phase will set in between the semimetal and
the featureless Mott phase. In Sec.\,\ref{sec: SO(3)}, we will
discuss one such scenario of the intermediate phase.

\subsection{$\SO(3)_\ell$ Symmetry Breaking Phase}\label{sec: SO(3)}

In the previous discussion of the semimetal to featureless Mott
transition, we focus on the scenario of a direct transition via
the SMG critical point. However, another possibility is that an
intermediate $\SO(3)_\ell$ SSB phase may set in, splitting the SMG
transition into two separate transitions. Starting from the
semimetal phase, the system can first develop a long-range order
of the $\O(3)$ vector $\langle\vect{M}\rangle\neq0$, gapping out
the physical fermions from the low-energy sector. Then the order
is destroyed upon the increasing interaction strength, which
restores the $\SO(3)_\ell$ symmetry in the featureless Mott phase.

\begin{figure}[htbp]
\begin{center}
\includegraphics[width=0.46\textwidth]{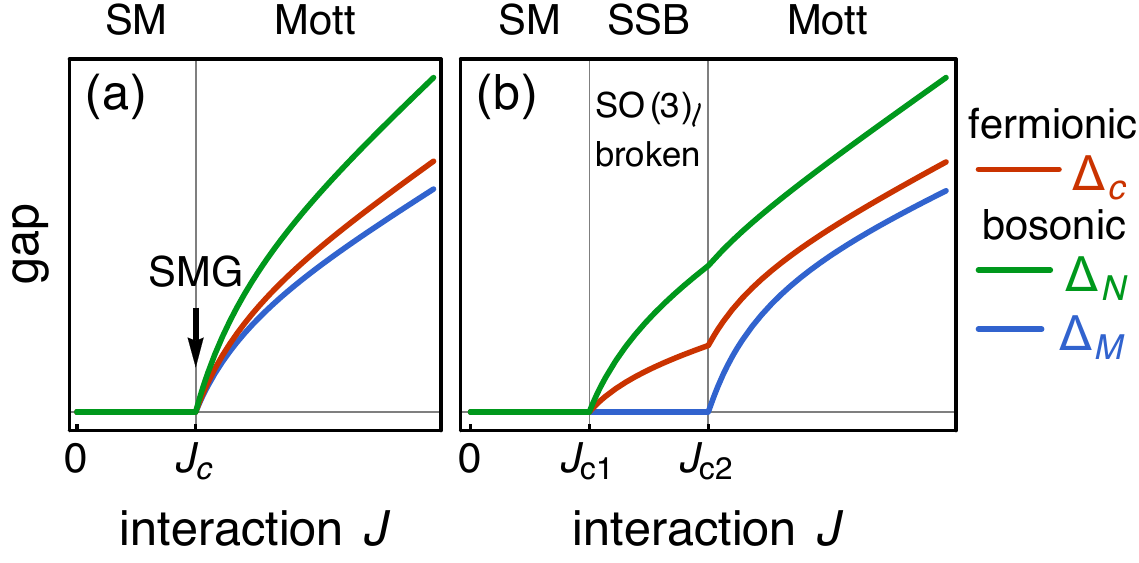}
\caption{Illustration of excitation gaps through the semimetal to
featureless Mott transition via (a) an SMG point, (b) an
intermediate $\SO(3)_\ell$ symmetry breaking phase. $J_c$ is an
SMG critical point, $J_{c1}$ is a Gross-Neveu critical point, and
$J_{c2}$ is an $\O(3)$ Wilson-Fisher critical point. In the strong
interaction $J\to\infty$ limit, we expect $\Delta_c\sim9J/8$,
$\Delta_N\sim3J/2$ and $\Delta_M\sim J$ to match the on-site
interaction spectrum listed in \tabref{tab: Hint spec}.}
\label{fig: gaps}
\end{center}
\end{figure}

These two scenarios can be distinguished from the different
behaviors of the excitation gaps as we tune the interaction. We
will focus on the following excitation gaps, defined via the
correlation functions
\eqs{\langle c_i^\dagger(\tau)c_j(0)\rangle&\sim e^{-\Delta_c \tau},\\
\langle \vect{N}_i(\tau)\cdot\vect{N}_j(0)\rangle&\sim e^{-\Delta_N \tau},\\
\langle \vect{M}_i(\tau)\cdot\vect{M}_j(0)\rangle&\sim
e^{-\Delta_M \tau},\\} where $\Delta_c$ the single-particle gap,
$\Delta_N$ is the $\O(4)$ vector gap and $\Delta_M$ is the $\O(3)$
vector gap. In the first scenario \figref{fig: gaps}(a), all
excitation gaps open up at the single SMG critical point. In the
second scenario  \figref{fig: gaps}(b), $\Delta_c$ and $\Delta_N$
first opens at a Gross-Neveu\cite{Gross:1974fc} critical point
$J_{c1}$, where the $\SO(3)_\ell$ symmetry is spontaneously
broken. Since the $\vect{N}$-vector boson excitation involves two
fermion excitations, so a gap $\Delta_N\simeq 2\Delta_c$ is
expected on the mean-field level.  Gapless Goldstone bosons of
$\vect{M}$ appear in the low-energy spectrum, so $\Delta_M$
remains zero in the $\SO(3)_\ell$ SSB phase. With stronger
interaction, the $\Delta_M$ gap eventually opens at the $\O(3)$
Wilson-Fisher\cite{Wilson:1972jx} critical point $J_{c2}$, where
the Goldstone modes of $\vect{M}$ are gapped.

In the parton field theory, the SMG critical point corresponds to
the case when the gap opening of the bosonic parton $\phi$ and the
spontaneous mass generation of the fermionic parton $\psi$ happen
at the same point. If the fermion mass generation happens before
the gapping of bosons, our theoretical framework will allow an
intermediate $\SO(3)_\ell$ SSB phase. In the parton language, the
Gross-Neveu transition $J_{c1}$ corresponds to the mass generation
of fermionic partons in the presence of the condensed bosonic
partons. However the parton description is not necessary in this
case because the condensed bosonic parton will Higgs out the gauge
field and make the fermionic parton equivalent to the physical
fermion, then the Gross-Neveu transition is just the conventional
mass generation for the physical fermion.

The $\O(3)$ Wilson-Fisher transition $J_{c2}$ turns out to be a
more interesting one, which is a transition of the bosonic parton
on the background of gapped fermionic partons. It is accessible
from the SMG theory \eqnref{eq: SMG} as
\eq{\scL_\text{WF}=\scL_\text{SMG}+m_\text{Mott} \tilde{M}^3,}
assuming the fermionic parton has developed an $\SU(2)$ gauge
triplet mass along the direction $\tilde{\vect{M}}\propto(0,0,1)$.
The mass $m_\text{Mott}$ gaps out the fermionic parton and Higgs
down the $\SU(2)$ gauge group to $\U(1)$, so the remaining
low-energy theory only contains bosonic partons coupled to the
$\U(1)$ gauge field
\eq{\scL_\text{WF}=\frac{1}{2}\big((\partial-\ii a^3\uptau^3-\ii
A_\ell^a\upmu^a)\upphi\big)^2+\frac{r}{2}\upphi^2+\frac{u}{4}
\upphi^4,} which is exactly the CP$^1$ field theory description of
the $\O(3)$ Wilson-Fisher critical point. The SSB-Mott transition
is driven by the bosonic mass term $r$. Since the physical fermion
excitation involves the excitations of both the fermionic and the
bosonic parton, so the gap opening of the bosonic parton will
create a kink in the single-particle gap $\Delta_c$ across the
transition $J_{c2}$, as illustrated in \figref{fig: gaps}(b).

\section{Summary}

In this work we have given an alternative description of the
bosonic topological transition (BTT), a transition between the
bosonic symmetry protected topological (BSPT) phase and the
featureless Mott phase, and demonstrated that how the BTT is
connected to another exotic quantum phase transition which we call the symmetry mass generation (SMG). Previously the BTT is described
by the $N_f = 2$ non-compact QED with an emergent $\O(4)$ symmetry in
the infrared, while in our work we show that the $N_f=4$ compact
QED is instable against a $\SU(4)$ to $\SO(4)$ breaking deformation, and
flows to a stable fixed point with $\O(4)=\SO(4)\rtimes\dsZ_2$
symmetry, which we identify with the infrared limit of the $N_f =
2$ non-compact QED.

The SMG is a direct transition between
the Dirac semimetal and the featureless Mott insulator, where the
Dirac fermions are gapped by their interaction without breaking
any symmetry. Conventionally, within Landau's paradigm, two
transitions, a Gross-Neveu followed by a Wilson-Fisher, are
expected between the semimetal and the featureless Mott phases. As
the two transitions merge into a single one, exotic quantum
criticality emerges beyond Landau's paradigm. We propose that the
SMG is a new type of deconfined quantum critical point (DQCP)
where the physical fermion is fractionalized into bosonic and
fermionic partons.

In particular, motivated by recent numerics, we studied a
$\SO(4)\times\SO(3)$ symmetric (2+1)D fermion model that exhibits
the SMG transition. At the critical point, the $\SO(3)$ symmetry
quantum number is carried by the bosonic parton and the $\SO(4)$
symmetry quantum number is carried by the fermionic parton. They
both couple to an emergent $\SU(2)$ gauge field as fundamental
representations. We propose that the SMG critical point can be described by the $\SU(2)$ QCD-Higgs theory,
where both the bosonic and fermionic partons become critical. Several key ingredients of the SMG theory are summarized as
follows. First, the theory must contain a Higgs field (e.g.\,the
bosonic parton), whose condensation can break the gauge structure
completely and bring the theory back to the semimetal phase.
Second, the $\dsZ_2^\scT$ symmetry acting on the fermionic parton
must be followed by a non-trivial gauge transformation, such that
a parton bilinear mass is allowed by the PSG to gap out the
fermionic partons in the featureless Mott phase. The parton
bilinear mass (e.g.\,the $\SU(2)$ triplet mass) must not be gauge
neutral, in order for the gauge transformation to come into play.
Finally, after the parton bilinear mass condensation, the
remaining unbroken gauge group (if not trivial) should be confined
by itself, such that a direct continuous SMG transition becomes
possible. These features are shared among the SMG theories with other symmetries as discussed in Ref.\,\onlinecite{Witten:2016yb,You:2017hy}.

\begin{acknowledgments}
We would like to acknowledge the helpful discussion with Chong
Wang, T. Senthil, Max Metlitski, Subir Sachdev. AV and YZY was
supported by a Simons Investigator grant. YCH is supported by a
postdoctoral fellowship from the Gordon and Betty Moore
Foundation, under the EPiQS initiative, GBMF4306, at Harvard
University. CX is supported by the David and Lucile Packard
Foundation and NSF Grant No. DMR-1151208.
\end{acknowledgments}

\appendix
\section{Majorana Basis}\label{sec: Maj basis}
\subsection{Lattice Model Basis}
In this section, we derive the low-energy effective theory from
the lattice model. Let us start from the Haldane model of
\emph{spinless} fermion on the honeycomb lattice:
\eq{H=-t\sum_{\langle ij\rangle}c_i^\dagger
c_j+\lambda\sum_{\langle\!\langle i j\rangle\!\rangle}\ii \nu_{ij}
c_i^\dagger c_j+\text{h.c.}.} Switching to the momentum space and
introducing $c_\vect{k}=(c_{\vect{k}A},c_{\vect{k}B})^\intercal$,
we have
\eqs{H&=\sum_{\vect{k}}c_\vect{k}^\dagger\mat{g(\vect{k})&f^*(\vect{k})\\f(\vect{k})&-g(\vect{k})}c_\vect{k},\\
f(\vect{k})&=-t\Big(e^{\ii k_y}+2e^{-\ii\frac{k_y}{2}}\cos\frac{\sqrt{3}k_x}{2}\Big),\\
g(\vect{k})&=-4\lambda\sin\frac{\sqrt{3}k_x}{2}\Big(\cos\frac{\sqrt{3}k_x}{2}-\cos\frac{3k_y}{2}\Big).}
Expand $f(\vect{k})$ and $g(\vect{k})$ around the momentum points
$K,K'=(\pm\frac{4\pi}{3\sqrt{3}},0)$, we get
\eq{\hspace{-6pt}\begin{array}{ll}&f(K+\vect{k})=v_F(k_x-\ii k_y),\\
&f(K'+\vect{k})=v_F(-k_x-\ii k_y),\end{array}
\begin{array}{ll}&g(K)=m_\text{IQH},\\
&g(K')=-m_\text{IQH},\end{array}} where $v_F=3t/2$ is the Fermi
velocity and $m_\text{IQH}=3\sqrt{3}\lambda$ is the integer
quantum Hall mass. In the following, we will set $v_F=1$ as the
energy unit. In the complex fermion basis (valley $\otimes$
sublattice) \eq{c=\smat{K\\K'}\otimes\smat{A\\B},} the low-energy
effective Hamiltonian (density) reads \eq{\label{eq: H
complex}\scH=c^\dagger(-\ii\partial_x\sigma^{31}+\ii\partial_y\sigma^{02}+m_\text{IQH}\sigma^{33})c.}
Throughout this appendix, we will follow the convention to denote
the tensor product of Pauli matrices as
\eq{\sigma^{abc\cdots}\equiv\sigma^a\otimes\sigma^b\otimes\sigma^c\otimes\cdots,}
where $a,b,c,\cdots=0,1,2,3$ are called the \emph{Pauli indices}.

Under translation $T_{1,2}: \vect{r}\to\vect{r}+\vect{a}_{1,2}$,
with $\vect{a}_{1,2}=(\sqrt{3}/2,\mp 3/2)$ according to
\figref{fig: honeycomb}. So the fermion transforms as \eq{T_{1,2}:
c\to e^{\ii\frac{2\pi}{3}\sigma^{30}}c.} The translation is
implemented as a three-fold rotation in the valley subspace. For
the chiral symmetry $\scS: c_i\to\scK(-)^ic_i^\dagger$, so
$c_{Q}\to\scK\sigma^{3}c_{Q}^\dagger$. Therefore, we conclude
\eq{\scS:c\to\scK\sigma^{03}c^\dagger.}

In the Majorana fermion basis (valley $\otimes$ sublattice
$\otimes$ particle-hole)
\eq{\upc=\smat{K\\K'}\otimes\smat{A\\B}\otimes\smat{\Re c\\\Im
c},} the Hamiltonian in \eqnref{eq: H complex} becomes
\eq{\label{eq: H Majorana}
\scH=\tfrac{1}{2}\upc^\intercal(-\ii\partial_x\sigma^{310}+\ii\partial_y\sigma^{022}+m_\text{IQH}\sigma^{332})\upc.}
We follow the convention of using the italic letter $c$ for
complex fermion and the upright letter $\upc$ for Majorana (real)
fermion. The symmetry actions can be written in the Majorana basis
as \eqs{\label{eq: symm Majorana}
T_{1,2}&:\upc\to e^{\ii\frac{2\pi}{3}\sigma^{302}}\upc,\\
\scS&:\upc\to\scK\sigma^{030}\upc.}

We introduce the spin and the layer degrees of freedom by
extending the Majorana basis to
\eq{\upc=\mat{K\\K'}\otimes\mat{A\\B}\otimes\mat{\uparrow\\\downarrow}\otimes\mat{1\\2}\\\otimes\mat{\Re
c\\\Im c}.} The Hamiltonian in \eqnref{eq: H Majorana} is also
extended to \eq{\label{eq: H
lattice}\scH=\tfrac{1}{2}\upc^\intercal(-\ii\partial_x\sigma^{31000}+\ii\partial_y\sigma^{02002}+m_\text{IQH}\sigma^{33002})\upc.}
The discrete symmetries is extended from \eqnref{eq: symm
Majorana} to \eqs{\label{eq: symm lattice}
T_{1,2}&:\upc\to e^{\ii\frac{2\pi}{3}\sigma^{30002}}\upc,\\
\scS&:\upc\to\scK\sigma^{03000}\upc.} 
The $\O(4)$ vector
$\vect{N}$ and the $\O(3)$ vector $\vect{M}$ can be written down following \eqnref{eq:
def NM}, \eqs{\label{eq: vector lattice}
\vect{N}&=\tfrac{1}{2}\upc^\intercal(\sigma^{03132},-\sigma^{10213},-\sigma^{10211},\sigma^{03230})\upc,\\
\vect{M}&=\tfrac{1}{2}\upc^\intercal(\sigma^{03012},\sigma^{03020},\sigma^{03332})\upc.}
As we can see, the rotation among these order parameters are
interwound with the valley and the sublattice degrees of freedom
in the lattice model basis.

\subsection{SMG Field Theory Basis}

To separate the action subspace of the discrete and the continuous
symmetries explicitly, we introduce the SMG field theory basis by
the following basis transformation from the lattice model basis,
\eq{\upc\to\mathsf{C}^{\sigma^{00300}}_{\sigma^{00030}}\mathsf{C}^{\sigma^{03000}}_{\ii\sigma^{00002}}\mathsf{C}^{\sigma^{01000}}_{\ii\sigma^{30002}}\mathsf{C}^{\sigma^{10000}}_{\ii\sigma^{00002}}\upc,}
where $\mathsf{C}^X_Y\equiv\tfrac{1}{2}(1+X+Y-XY)$ represents a
controlled gate. Under the transformation, the Hamiltonian in
\eqnref{eq: H lattice} becomes \eq{\label{eq: H QFT}
\scH=\tfrac{1}{2}\upc^\intercal(-\ii\partial_x\sigma^{03000}+\ii\partial_y\sigma^{01000}+m_\text{IQH}\sigma^{02000})\upc.}
The symmetry actions in \eqnref{eq: symm lattice} become
\eqs{\label{eq: symm QFT}
T_{1,2}&:\upc\to e^{\ii\frac{2\pi}{3}\sigma^{20000}}\upc,\\
\scS&:\upc\to\scK\sigma^{22000}\upc,} which leads to
Eqs.\,(\ref{eq: T12}, \ref{eq: Z2S}) in the complex fermion basis.
Note that the $C_6$ transform can be equivalently written as
$C_6:\upc\to\ii\sigma^{30002}e^{\ii\frac{\pi}{6}\sigma^{02000}}\upc$.
The generator $\sigma^{20000}$ of the translation indicates that
the valley basis has been transformed to
$\ket{K_\pm}=(\ket{K}\pm\ii\ket{K'})/\sqrt{2}$. The fermion
bilinear order parameters in \eqnref{eq: vector lattice} become
\eqs{\label{eq: vector QFT}
\vect{N}&=\tfrac{1}{2}\upc^\intercal(\sigma^{22102},\sigma^{22123},\sigma^{22121},\sigma^{22200})\upc,\\
\vect{M}&=\tfrac{1}{2}\upc^\intercal(\sigma^{22312},\sigma^{22320},\sigma^{22332})\upc.}
The continuous symmetry $\SO(4)\times\SO(3)_\ell$ acts only in the
spin $\otimes$ layer $\otimes$ particle-hole subspace. The
representation of their generators in the Majorana basis can be
derived from the commutator of the vectors $\vect{N}$ and
$\vect{M}$ in the same representation, which are concluded in
\tabref{tab: generators}
\begin{table}[htbp]
\caption{Representations of the $\SO(4)\times\SO(3)_\ell$
generators in the spin $\otimes$ layer $\otimes$ particle-hole
subspace. The generator at row-$a$ column-$b$ is given by
$\Gamma_{ab}\sim\ii[\Gamma_a,\Gamma_b]$ (where the sign and
pre-factors are omitted).} \vspace{5pt}
\begin{minipage}[t]{0.5\linewidth}
$\SO(4)$ generators\\\vspace{5pt}
\begin{tabular}{c|cccc}
 & $\sigma^{102}$ & $\sigma^{123}$ & $\sigma^{121}$ & $\sigma^{200}$\\
\hline
$\sigma^{102}$ & & $\sigma^{021}$ & $\sigma^{023}$ & $\sigma^{302}$\\
$\sigma^{123}$ & & & $\sigma^{002}$ & $\sigma^{323}$\\
$\sigma^{121}$ & & & & $\sigma^{321}$\\
$\sigma^{200}$ & & & &
\end{tabular}
\end{minipage}
\begin{minipage}[t]{0.4\linewidth}
$\SO(3)_\ell$ generators\\\vspace{5pt}
\begin{tabular}{c|ccc}
 & $\sigma^{312}$ & $\sigma^{320}$ & $\sigma^{332}$\\
\hline
$\sigma^{312}$ & & $\sigma^{032}$ & $\sigma^{020}$\\
$\sigma^{320}$ & & & $\sigma^{012}$\\
$\sigma^{332}$ & & &
\end{tabular}
\end{minipage}
\label{tab: generators}
\end{table}
From \tabref{tab: generators} we can see the $\SO(4)$ generators
splits to those of the $\SU(2)_\uparrow\times\SU(2)_\downarrow$
group. In each spin sector, the $\SU(2)_\sigma$ group acts only in
the layer $\otimes$ particle-hole subspace, whose generators are
represented as
\eq{\vect{\upmu}=\sigma^{000}\otimes(\sigma^{23},\sigma^{21},\sigma^{02}).}
The $\SO(3)_\ell\simeq\SU(2)_\ell$ group also acts only in the
layer $\otimes$ particle-hole subspace, whose generators are
represented as
\eq{\vect{\uptau}=\sigma^{000}\otimes(\sigma^{12},\sigma^{20},\sigma^{32}).}
So the symmetry charge are indeed given by \eqnref{eq: symm
charges}. For the consistency in the context, here we have filled
in the omitted identity operator $\sigma^{000}$ in the valley
$\otimes$ sublattice $\otimes$ spin subspace.

In the SMG field theory basis, there are only six mass terms of
the form $\upc^\intercal\mathsf{M}\upc$ that commute with all the
$\SO(4)\times\SO(3)_\ell$ generators:
\eqs{\mathsf{M}=\;&\sigma^{02000},\sigma^{12000},\sigma^{32000},\\
&\sigma^{02300},\sigma^{12300},\sigma^{32300}.} In the first line,
the first one ($\sigma^{02000}$) is the IQH mass. The following
two ($\sigma^{12000}$ and $\sigma^{32000}$) are the Kekul\'e
masses (dimerization according to the Kekul\'e pattern) which
breaks the translation symmetry, as they do not commute with the
translation generator $\sigma^{20000}$ given in \eqnref{eq: symm
QFT}. The second line is the spin-dependent version of the first
line. For example, the first one $\sigma^{02300}$ corresponds to
the QSH mass. Therefore if we consider translation invariant
masses, we are left with the IQH and QSH masses only. It is easy
to see that these two masses are further ruled out by the
reflection $\sigma_h$, the time-reversal $\scT$ and the chiral
$\scS$ symmetries, given their definitions in \eqnref{eq: symm
QFT}.

The fermionic parton can be written in the same field theory basis
as the physical fermion by replacing $\upc\to\uppsi$ in all
equations. In the featureless Mott phase, we expect a $\vect{M}$
mass (for example $M^3$) will be generated for the fermionic
partons. From \eqnref{eq: symm QFT}, we can see $M^3$ changes sign
under $\scS$ transformation. So the chiral symmetry should act
projectively (i.e.\,should be followed by the gauge transformation
$\uppsi\to\ii\sigma^{00020}\uppsi$ to revert the sign change of
the mean-field mass $M^3$). \eqs{\label{eq: symm PSG}
T_{1,2}&:\uppsi\to e^{\ii\frac{2\pi}{3}\sigma^{20000}}\uppsi,\\
\scS&:\uppsi\to\scK\ii\sigma^{22020}\uppsi.}

For the fermionic partons, the $\SU(2)_\ell$ symmetry is promoted
to the $\SU(2)$ gauge group. As the fermionic partons are gapped
out by the mass $M\uppsi^\intercal\sigma^{22332}\uppsi$, the
$\SU(2)$ gauge group will be Higgs down to its $\U(1)$ subgroup.
The $\U(1)$ gauge flux $\dd a^3\uptau^3$ can acquire a quantum
number due to the Hall-like response. The quantum number is
determined by evaluating the following product
\eq{\mathsf{q}=\gamma^0M^3\uptau^3=\sigma^{02000}\sigma^{22332}\sigma^{00032}=\sigma^{20300}.}
The transformation $\uppsi\to e^{\ii\theta\mathsf{q}}\uppsi$
generated by this quantum number correspond to a spin-dependent
translation (recall that the translation generator is
$\sigma^{20000}$ in \eqnref{eq: symm PSG} while
$\sigma^{00300}=(-)^\sigma$ is the sign of the spin and
$\mathsf{q}$ is just the product of them). Since the
spin-dependent translation is not a symmetry at the UV scale, the
$\U(1)$ gauge field $a^3$ will remain compact.

\subsection{BTT Field Theory Basis}

Near the bosonic topological transition, the masses $M$ and
$m_\text{QSH}$ competes in the fermionic parton sector,
\eqs{\scH=\tfrac{1}{2}\uppsi^\intercal(&-\ii\partial_x\sigma^{03000}+\ii\partial_y\sigma^{01000}\\
&+M\sigma^{22332}+m_\text{QSH}\sigma^{02300})\uppsi.} At the
transition $|M|=|m_\text{QSH}|$, half of the fermions $\uppsi$
will be gapped out. To focus on the remaining gapless fermions, we
wish to explicitly separate gapped and gapless subspaces. This can
be done by further applying the following transformation to the
field theory basis,
\eq{\uppsi\to\mathsf{C}^{\sigma^{10030}}_{-\ii\sigma^{00002}}\mathsf{C}^{\sigma^{00030}}_{\sigma^{00003}}\uppsi,}
under which the Hamiltonian becomes \eqs{\label{eq: H new basis}
\scH=\tfrac{1}{2}\uppsi^\intercal(&-\ii\partial_x\sigma^{03000}+\ii\partial_y\sigma^{01000}\\
&+M\sigma^{32300}+m_\text{QSH}\sigma^{02300})\uppsi.} In this
basis, the masses $M$ and $m_\text{QSH}$ only differed by the
factor of $\sigma^{30000}$, then the gapped and gapless subspace
are clearly separated by \eq{\label{eq: UV-IR
indicator}\sigma^{30000}\uppsi=\left\{\begin{array}{ll} +\uppsi &
\text{gapped,}\\-\uppsi & \text{gapless.}\end{array}\right.} The
PSG transformations in \eqnref{eq: symm PSG} becomes
\eqs{\label{eq: symm PSG new basis}
T_{1,2}&:\uppsi\to e^{\ii\frac{2\pi}{3}\sigma^{30002}}\uppsi,\\
\scS&:\uppsi\to\scK\ii\sigma^{22023}\uppsi.} The translation
generator is transformed back to $\sigma^{30002}$, implying that
the valley subspace is represented in the $\ket{K}$, $\ket{K'}$
basis again (nominally). So \eqnref{eq: UV-IR indicator} simply
suggest that the fermions $\uppsi_K$ from the $K$-valley are
gapped and those from the $K'$-valley $\uppsi_{K'}$ remain
gapless. Nevertheless, the statement should not be taken too
seriously, since the basis transformation has mixed the valley and
gauge charge together. In the new basis, the $\vect{N}$ and
$\vect{M}$ vectors are represented as \eqs{
\vect{N}[\uppsi]&=\tfrac{1}{2}\uppsi^\intercal(\sigma^{32130},\sigma^{32110},\sigma^{32122},\sigma^{32202})\uppsi,\\
\vect{M}[\uppsi]&=\tfrac{1}{2}\uppsi^\intercal(\sigma^{22321},\sigma^{22323},\sigma^{32300})\uppsi,}
and the $\SU(2)_\sigma$ and $\SU(2)_\ell$ generators are represented as
\eqs{
\vect{\upmu}&=\sigma^{000}\otimes(\sigma^{12},\sigma^{20},\sigma^{32}),\\
\vect{\uptau}&=(\sigma^{10023},\sigma^{10021},\sigma^{00002}).}

Now we project to the low-energy subspace of
$\sigma^{30000}\uppsi=-\uppsi$, this amounts to removing the first
Pauli index in all Pauli operators. Off-diagonal operators
$\sigma^{1\cdots},\sigma^{2\cdots}$ will not survive the
projections (i.e.\,they can not be represented in the low-energy
subspace). The Hamiltonian \eqnref{eq: H new basis} is reduced to
\eq{\scH=\tfrac{1}{2}\uppsi^\intercal(-\ii\partial_x\sigma^{3000}+\ii\partial_y\sigma^{1000}
+m'_\text{QSH}\sigma^{2300})\uppsi,} where
$m'_\text{QSH}=m_\text{QSH}-M$ is the reduced QSH mass. This
low-energy subspace basis is the BTT field theory basis. From
\eqnref{eq: symm PSG new basis}, we can see that the chiral
symmetry $\scS$ does not survive the the projection, because the
QSH mass $m_\text{QSH}$ has explicitly broken the symmetry. The
$\SO(4)\simeq\SU(2)_\uparrow\times\SU(2)_\downarrow$ symmetry is
preserved. The $\O(4)$ vector is projected to
\eq{\vect{N}=-\tfrac{1}{2}\uppsi^\intercal(\sigma^{2130},\sigma^{2110},\sigma^{2122},\sigma^{2202})\uppsi.}
The $\SU(2)_\sigma$ generators becomes
$\vect{\upmu}=\sigma^{00}\otimes(\sigma^{12},\sigma^{20},\sigma^{32}).$
The $\SO(3)_\ell\simeq\SU(2)_\ell$ symmetry is now gauged for the
fermionic partons. The mass $M$ explicitly Higgs down the gauge
group from $\SU(2)$ to $\U(1)$ and the remaining $\U(1)$ gauge
charge is $\uptau^3=\sigma^{00002}$. So in the BTT field theory
basis, the $\U(1)$ gauge transformation is simply the phase
rotation $\psi\to e^{\ii\theta}\psi$ of the complex fermion
$\psi$.

\bibliography{SMG}
\bibliographystyle{apsrev}

\end{document}